\documentclass[12pt]{iopart}
\usepackage{graphicx}
\usepackage[breaklinks=true,colorlinks=true,linkcolor=blue,urlcolor=blue,citecolor=blue]{hyperref}
\usepackage{subfig}

\usepackage{cite}
\usepackage{soul}  

\begin{document}

\title[Cross-Calibration and First V-ECE Measurement of Electron Energy Distribution in TCV]{Cross-Calibration and First Vertical ECE Measurement of Electron Energy Distribution in the TCV Tokamak}

\author{A Tema Biwole$^1$\footnote{Present address:
MIT - Plasma Science and Fusion Center, Cambridge, Massachusetts 02139, USA}, L Porte$^1$, A Fasoli$^1$, L Figini$^2$, J Decker$^1$, M Hoppe$^3$, J Cazabonne$^4$, L Votta$^3$, A Simonetto$^2$, S  Coda$^4$, and the TCV Team$^5$ } 

\address{$^1$ Ecole Polytechnique Fédérale de Lausanne (EPFL), Swiss Plasma Center (SPC), CH-1015 Lausanne, Switzerland}
\address{$^2$ Istituto per la Scienza e Tecnologia dei plasmi, Via Roberto Cozzi 53 - 20125 Milano (MI)}
\address{$^3$ KTH Royal Institute of Technology, Department of Electrical Engineering, Stockholm, Sweden }
\address{$^4$ Commissariat à L’énergie Atomique, CEA-IRFM F-13108 Saint-Paul-les-Durance, France}
\address{$^5$ See authors list in  H. Reimerdes et al. Nucl. Fusion 62, 042018 (2022).}

\ead{biwole@mit.edu}
\vspace{10pt}
\begin{indented}
\item[]\today
\end{indented}

\begin{abstract}

This paper describes the first Vertical Electron Cyclotron Emission (V-ECE) measurement of non-thermal electron distributions in the \textit{Tokamak à Configuration Variable}, TCV. These measurements were conducted in runaway electron scenarios and in the presence of Electron Cyclotron Current Drive. Measured intensities of linearly polarised X- and O-mode radiation from fast  electrons allow the analysis of the energy distribution. The measurements were made possible through the creation of an operational regime for the diagnostic that is free of thermal background radiation, in relaxed electron density operations. This operational regime notably enables the cross-calibration of the diagnostic system, relying on thermal plasma measurements and modeling with the ray-tracing code SPECE.

\end{abstract}

%
\vspace{2pc}
\\
\\
\vspace{2pc}
%
\submitto{\PPCF}
%
%
%

\section{Introduction}
\label{intro}
Electron Cyclotron Emission (ECE) is a well-established technique for measuring electron temperature in Tokamaks \cite{Hutchinson2002,harfuss_mm_ch4}. The conditions for such measurement generally include a thermal plasma, i.e., a plasma in which the electron distribution can be assumed Maxwellian, and optically thick conditions for the electron cyclotron wave. For temperature measurement, when these conditions are fulfilled, ECE configurations are usually chosen to have a horizontal line of sight, viewing the plasma perpendicularly from the Low Field Side (LFS ECE) or from the High Field Side (HFS ECE). In these horizontal configurations, ECE measurements at the $n$ harmonic take advantage of the $1/R$ dependence of the electron cyclotron frequency \cite{Bornatici_1983}, $\omega_{n\mathrm{ece}}\sim n eB(R)/m_{\mathrm{e}}$, to infer the radial profile of the electron temperature, $T_{\mathrm{e}}(R)$.
Electron Cyclotron Emission has also been used to diagnose plasmas in which the electron distribution has departed from a Maxwellian distribution. For the diagnosis of non-thermal electrons, the Vertical ECE configuration (V-ECE), with $R$ constant along the line of sight, has proven to have the advantage of a more straightforward discrimination of the electron energy according to the radiation frequency:
\begin{equation}
    \omega \sim n eB/ \gamma m_{\mathrm{e0}},
    \label{eq:one-to-one}
\end{equation}
where $\gamma$ is the relativistic factor and $m_{\mathrm{e0}}$ is the electron rest mass ($\gamma m_{\mathrm{e0}} = m_{\mathrm{e}}$).

The work presented in this paper, on the analysis of the electron energy distribution using V-ECE measurements, builds on previous works on Alcator C \cite{Kato1986a, Hutchinson1986a, Hutchinson1986, kato_phdthesis_1986, Kato_1986, Kato1987}, TEXT-U \cite{Giruzzi1996, Roberts_1995}, WT-3 \cite{Ide1989}, DIII-D \cite{James1988, janz_phdthesis_1992}, and PLT \cite{luce_phdthesis_1987, Luce1988, Luce1987} Tokamaks, which attempted to leverage the ideal one-to-one correspondence between measured frequency and electron energy in Equation \ref{eq:one-to-one}, by minimizing or mitigating issues such as harmonic overlap and multiple wall reflections. A method, described in Reference \cite{Michelot_1996}, was also attempted to infer electron temperature from V-ECE measurements.
In the following sections, we will discuss the experimental arrangement that enables the measurement of non-thermal electrons using V-ECE on TCV (Section \ref{setup}). The calibration of the diagnostic, which relies on V-ECE modeling and measurements free of background radiation, will be presented in Section \ref{calib}. V-ECE measurements of fast electron radiation in runaway electron scenarios and Electron Cyclotron Current Drive (ECCD) will be discussed in Section \ref{meas4b}. The analysis of the electron energy distribution based on these measurements is presented in Section \ref{meas4c_disc_edf}, followed by the general conclusion of the paper. All physical quantities in this paper are expressed using the International System of Units (SI). The angular frequency $\omega$ in rad/s will be preferred for equations, while the frequency $f$ in Hz will be used in quantitative assessments for practical reasons.
\section{Experimental Setup}
\label{setup}
\begin{figure}[ht!]
\centering    
\subfloat[]
{\label{fig:vece_optics}%
\includegraphics[width=0.3\columnwidth]{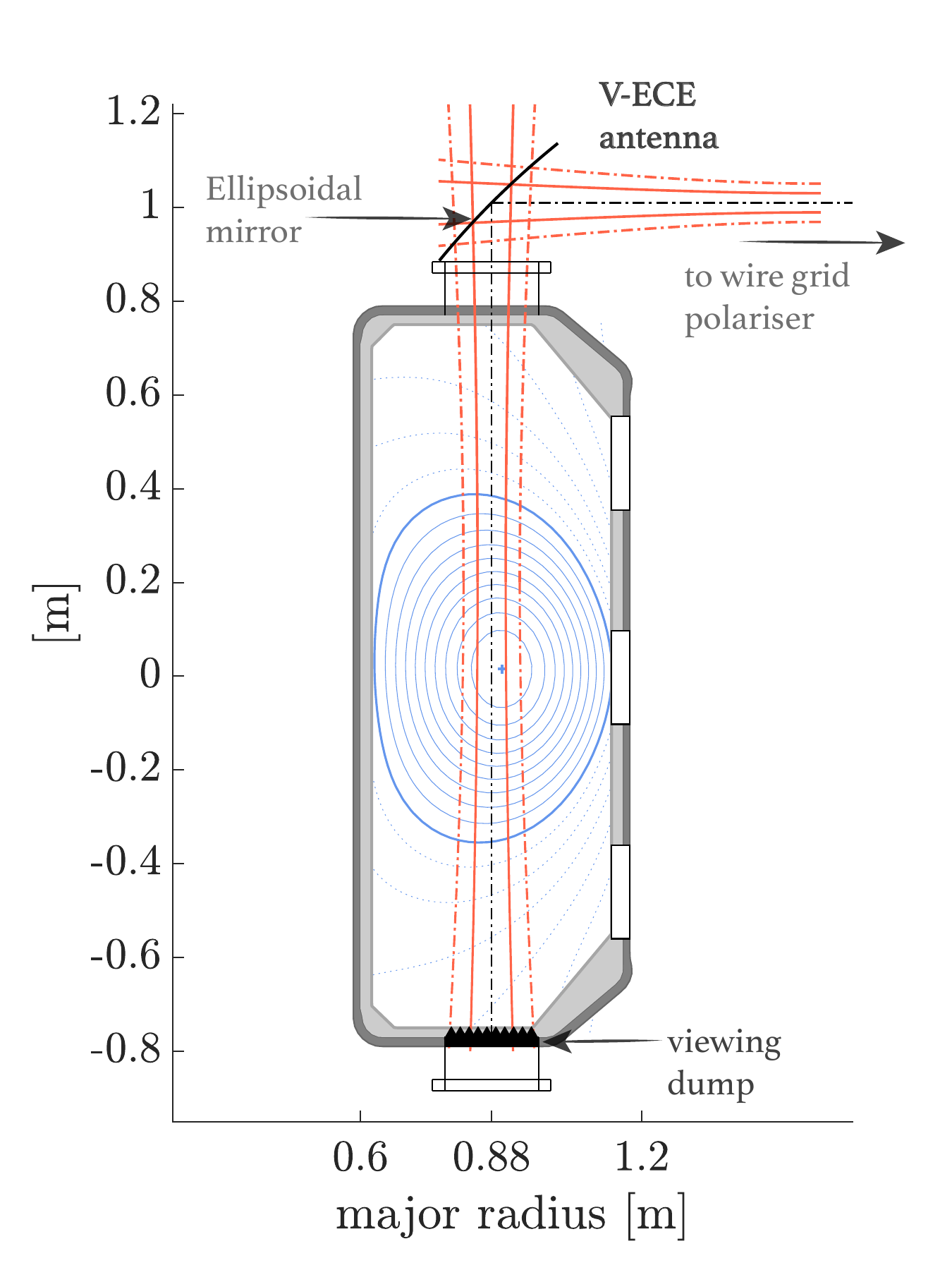}} \quad
\subfloat[]
{\label{fig:harmonics}%
\includegraphics[width=0.54\columnwidth]{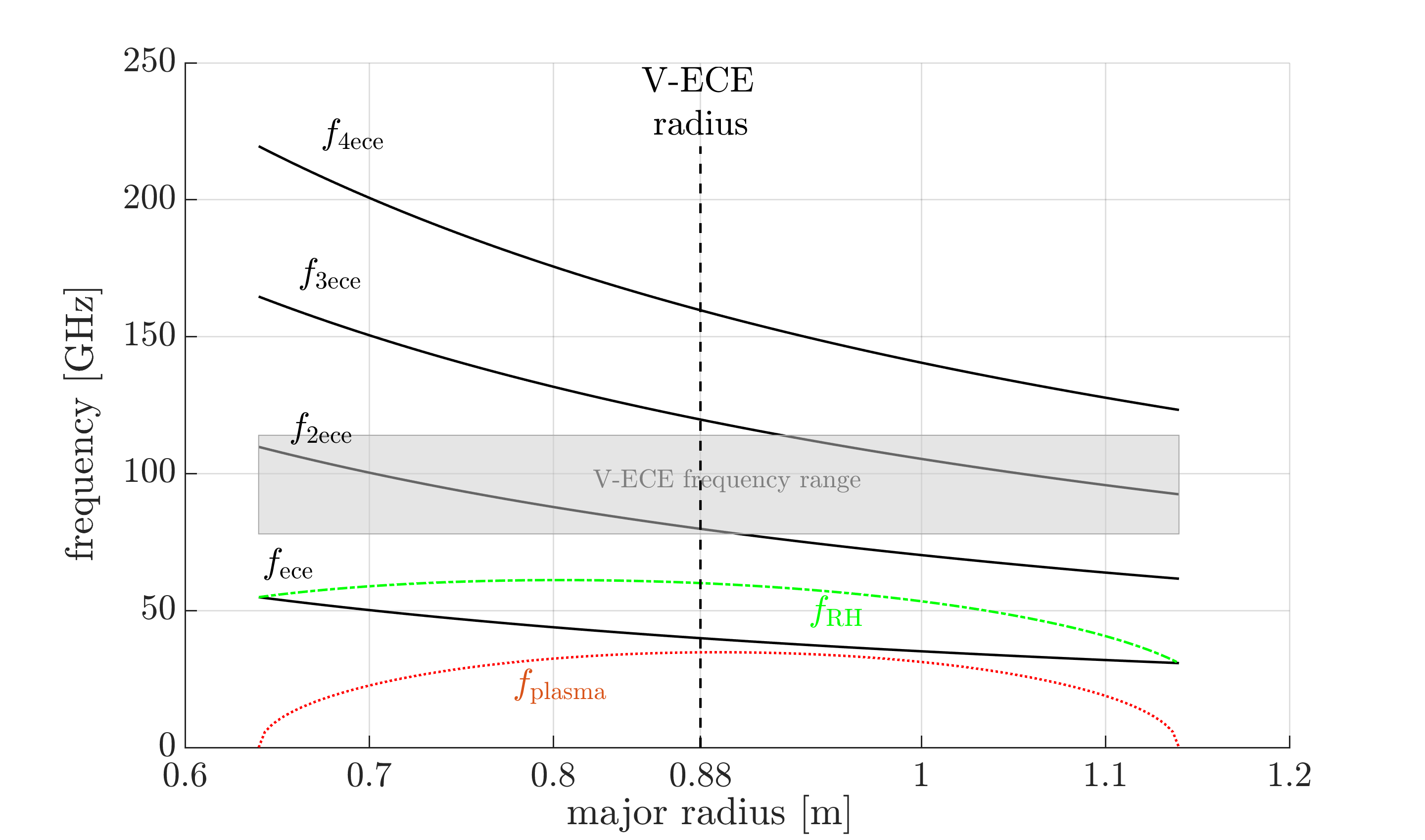}} 
\caption[\ref{fig:vece_optics}V-ECE optics on TCV. The line of sight, which terminates on a beam dump (dark component below the machine floor) forms a Gaussian beam of radius $\sim$ 3 cm near the mid-plane. The radiation is reflected on an ellipsoidal mirror, split into X- and O- polarisations before detection in a heterodyne radiometer.\ref{fig:harmonics} V-ECE frequency range,  ECE harmonic frequencies ($f_{n\mathrm{ece}}$) and Right-Hand cut-off frequency ($f_{RH}$) on TCV with a toroidal magnetic field $B_0$=1.42 T at the vacuum vessel centre $R_0$=0.88 m.  The represented plasma frequency is computed with a central electron density of 1.5 $\times 10^{19}$ m$^{-3}$.]{\ref{fig:vece_optics} V-ECE optics on TCV. The line of sight, which terminates on a beam dump (dark component below the machine floor) forms a Gaussian beam of radius $\sim$ 3 cm near the mid-plane. The radiation is reflected on an ellipsoidal mirror, split into X- and O- polarisations before detection in a heterodyne radiometer. \ref{fig:harmonics} V-ECE frequency range, ECE harmonic frequencies ($f_{n\mathrm{ece}}$), and Right-Hand cut-off frequency ($f_{RH}$) on TCV with a toroidal magnetic field $B_0$=1.42 T at the vacuum vessel centre $R_0$=0.88 m. The represented plasma frequency is computed with a central electron density of 1.5 $\times 10^{19}$ m$^{-3}$.}
\label{fig:setup}
\end{figure}

The measurements discussed in this paper are carried out using the V-ECE diagnostic\cite{Biwole2023_vece}, installed on the TCV Tokamak \cite{Reimerdes_2022}.
The diagnostic is equipped with a set of heterodyne radiometers measuring in the frequency range $78-114$ GHz, with temporal resolution in the order of $10 \mu$s and bandwidth $\sim 750$ MHz.
The radiometer, described in detail in \cite{TemaBiwole2023}, is composed of three main stages: a Radio Frequency (RF) stage, an Intermediate Frequency (IF) stage, and a video stage. The incoming radiation from the plasma is split into the X- and O- polarisation (X- and O-mode) using a wire grid polariser, then downshifted to lower frequencies in the RF stage. This frequency conversion occurs in a mixer, which is a diode with a non-linear I/V characteristic. The lower frequency signals are amplified in the IF stage, rectified by a square-law diode detector, and further processed in the video stage, where the signals are amplified and low-pass filtered down to 100 kHz for acquisition. The optics of the diagnostic is arranged such that the vertical antenna pattern forms a Gaussian beam of waist radius $\sim 3$ cm near the vessel mid-plane. In vacuum, the antenna pattern terminates, at the bottom of the machine, on a highly absorbing viewing dump \cite{Biwole2021_dump}, see Figure \ref{fig:vece_optics}. During plasma discharges, the antenna pattern can be shifted away from the viewing dump, due to refraction, at a threshold electron density close to $1 \times 10^{19} \mathrm{m}^{-3}$\cite{TemaBiwole2023}. The diagnostic measures fast electron emission at downshifted third or fourth harmonic emission as shown in Figure \ref{fig:harmonics}. The electron energy that can be measured at each frequency is calculated in\cite{Biwole2023_vece}. It is calculated that the maximum energy the diagnostic can measure without harmonic overlap is in the order of $\sim 250 $ keV. The ECCD experiments discussed in this paper exploit the  X2 gyrotron from the ECH, Electron Cyclotron Heating system on TCV\cite{Paley2009}. The gyrotron produces wave at $82.7$ GHz and  nominal power $\sim 680$ kW. The gyrotron is connected to a launcher which direct the beam into the plasma. Current drive is achieved by imparting a parallel component, with respect to the plasma magnetic field, to the ECH wave vector. The location of the beam power deposition along the radial coordinate is generally controlled by the chosen value of magnetic field strength. The toroidal and poloidal directions of the beam $($in the reference frame of the plasma $)$ are controlled by the launcher angles, so-called  launcher toroidal angle,  $\phi_{\mathrm{L}}$, and poloidal angle, $\theta_{\mathrm{L}}$ that can can be varied during discharges. 
\newpage

\section{Calibration}
\label{calib}

\subsection{Background}
\label{background}
The calibration of ECE diagnostics generally consists in determining, with the highest accuracy, the spectral response of the  system to a given radiation power. Traditionally, two approaches have been retained for  ECE calibration: (1) the use of sources of known characteristics as substitutes for the plasma and (2) the cross-calibration on a different, absolutely calibrated instrument.
In the first approach,  hot and cold sources  are positioned within the vacuum chamber to irradiate the ECE antenna\cite{Costley1974,COSTLEY2009}. The approach has been successfully employed since the 1970s in the vast majority of ECE systems in magnetic confinement devices, as demonstrated, for example, in Reference \cite{_newBuratti}. A limitation of this approach in the past was that the calibration needed to be conducted inside the vacuum vessel, necessarily during a machine vent. The system's response, determined during a hot/cold source calibration, was typically  used for several months, or even years, until the next calibration was feasible. This raised the question of the calibration's stability in relation to variations in experimental conditions, which could occur on a  daily basis. The limitations of the calibration with a reference source inside the vacuum vessel have been addressed in previous works, such as in Reference \cite{BURATTI1993533} by using external replica systems of ECE antennas, an application which is believed to be useful for ITER.
The latter calibration approach exploits the absolute calibration of a different instrument within the same Tokamak, to obtain the response function of an ECE  system. The ECE system can  be cross-calibrated on another ECE system as in JET \cite{Schmuck_2012}, or can be cross-calibrated on a different diagnostic which measures the same plasma parameters such as Thomson Scattering, inheriting all the uncertainties from the absolute calibration. This calibration method relies on the assumption that both diagnostics measure the same electron temperature. For ECE diagnostic systems,  the measurement of the electron temperature requires that the plasma is optically thick and  that the electron distribution is  Maxwellian. 
When these conditions are fulfilled, successful cross-calibration of ECE systems with Thomson Scattering (TS) can be achieved, as demonstrated over the years, on TCV for example
 \cite{Blanchard2002,blanchard_phdthesis_2002, klimanov_phdthesis_2005, Klimanov2005, Fontana_2017}. Historically, ECE and Thomson Scattering cross-calibration have used ohmic plasma discharges, which provide the best experimental plasma conditions allowing the assumption of a Maxwellian electron distribution. In other experimental conditions, such as those with NBI or ICRH heating, discrepancies between ECE and Thomson Scattering can occur, and modeling of the electron distribution is required \cite{de_la_Luna_2003,Fontana2023}. For a vertical ECE configuration, a direct cross-calibration with Thomson Scattering cannot be achieved for two main reasons. First, even in an optically thick plasma, the radiation intensity measured with a vertical line of sight cannot be straightforwardly associated with a local electron temperature, since the ECE resonance is not well localized in the vertical direction. Second, the plasma is not usually optically thick for a vertical ECE line of sight above a certain frequency (approximately 86 GHz on TCV). At these higher, optically thin frequencies, an accurate estimation of the absorption coefficient and, consequently, the radiation intensity needs to be performed in order to use the plasma emission for calibration.
Examples of calculations of radiation intensity in optically thin conditions have been reported in previous works, such as in DIII-D for the measurement of the electron temperature from the optically gray third harmonic \cite{Austin_1996}, and in JET for calibrating the optically thin O1 (first harmonic O-mode) radiation for temperature measurement in the inboard (High Field Side) pedestal \cite{Barrera_2010}. In both cases, a horizontal lines of sight was employed.
\subsection{Theory}
\label{sec:theo}
The ECE spectral intensity $I_{\omega}$ \cite{bekefi}, for a radiation path in the plasma of total length $L$ is expressed as 
\begin{equation}
     I_{\omega} = I_{\omega}^{\mathrm{inc}} e^{-\tau_0} + \frac{\omega^2}{8\pi^3c^2} \int_{0}^{L} T_{\mathrm{e}}(s) \; \alpha_{\omega}(s)\; e^{-\tau (s)} \;ds,
     \label{eq:93}  
\end{equation}
 where the coordinate $s$ measures the ray path along the plasma and $\tau_0$ is the total plasma optical depth. The absorption coefficient, $\alpha_{\omega}$, represents the rate of absorption of the radiation per unit path length while $T_{\mathrm{e}}(s)$ stands for the temperature of the radiating electrons along the ray path. The first term in Equation \ref{eq:93} represents the contribution of the incident radiation to the intensity, for example that of the thermal background radiation. It vanishes in case of no incident radiation, $I_{\omega}^{\mathrm{inc}} = 0 $,  or optical thick conditions $\tau_0 \gg 1$. In case of optically thin plasmas, $\tau_0 < 1$, an estimation of the contribution from multiple wall reflection is needed to assess $I_{\omega}^{\mathrm{inc}}$. That contribution can be neglected if the line of sight  avoids radiation from multiple wall reflections, with a viewing dump as an example.  
If the electron distribution is  Maxwellian and the plasma is optically thick ($\tau_0 \gg 1$), the ECE radiation intensity is simply proportional to a local electron temperature and straight cross-calibration against, say, Thomson Scattering temperature  can be achieved. In optically thin conditions and  vertical-viewing line of sight, the intensity, $I_{\omega}$ needs to be accurately estimated, by calculating the absorption coefficient \cite{Bornatici1982,Bornatici_1983} along the ray trajectory, e.g. with the ray-tracing code SPECE \cite{Farina_2008}, which solves the fully relativistic dispersion relation for EC waves. Experimentally, it is necessary to avoid incident radiation from multiple wall reflections for the accuracy of the calibration. 

\subsection{Measurement}

\begin{figure}[ht!] 
\centering 
\includegraphics[width=1.0\columnwidth]{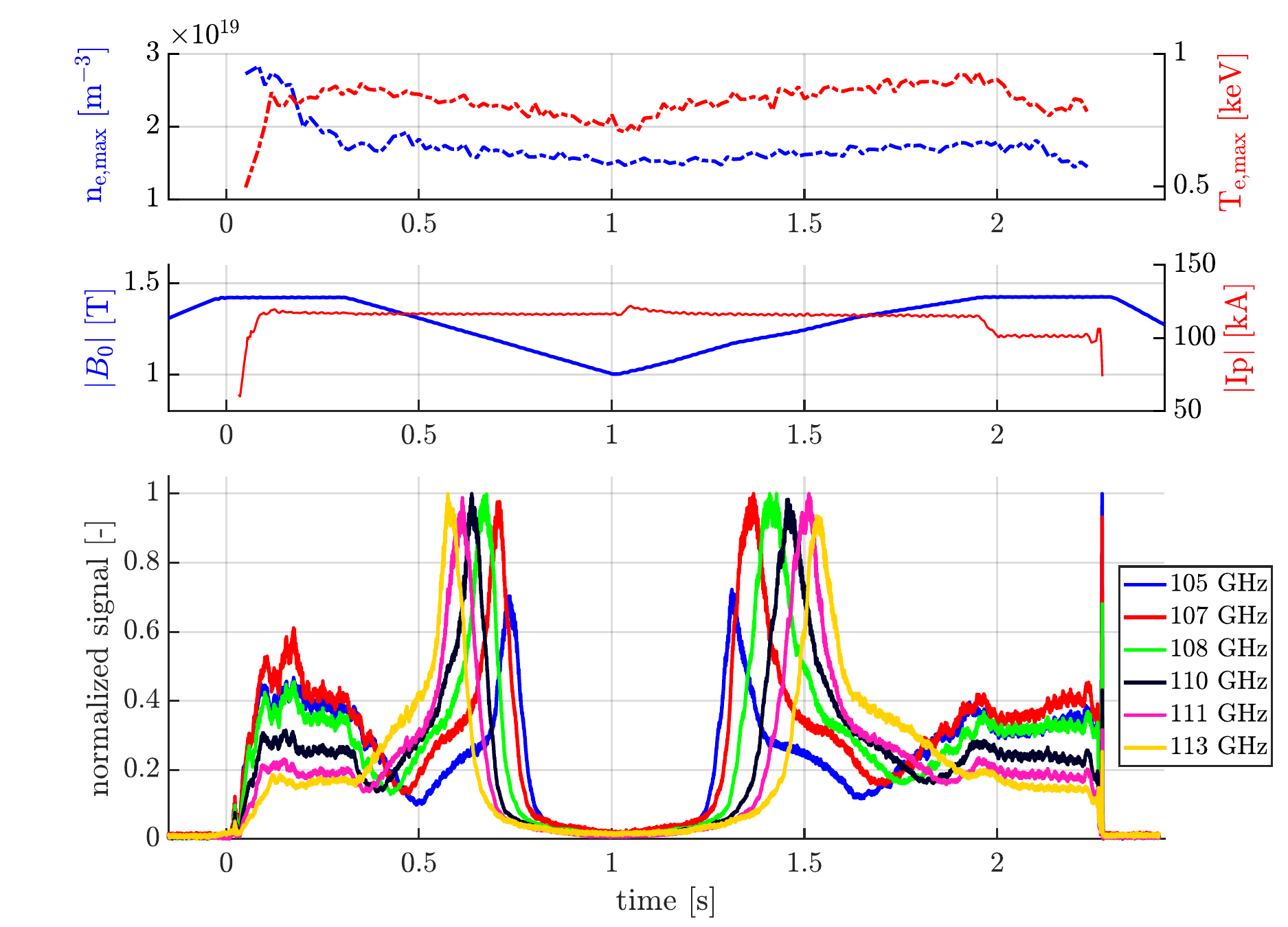} 
\caption[Measurement in the frequency range $104-114$ GHz during the calibration discharge $\#67946$ with two field ramps and flat plasma current.]{ Measurement in the frequency range $104-114$ GHz during the calibration discharge $\#67946$ with two field ramps and flat plasma current.}
\label{fig:radio2_67946_overview} 
\end{figure}

\begin{figure}[ht!] 
\centering 
\includegraphics[width=1.\columnwidth]{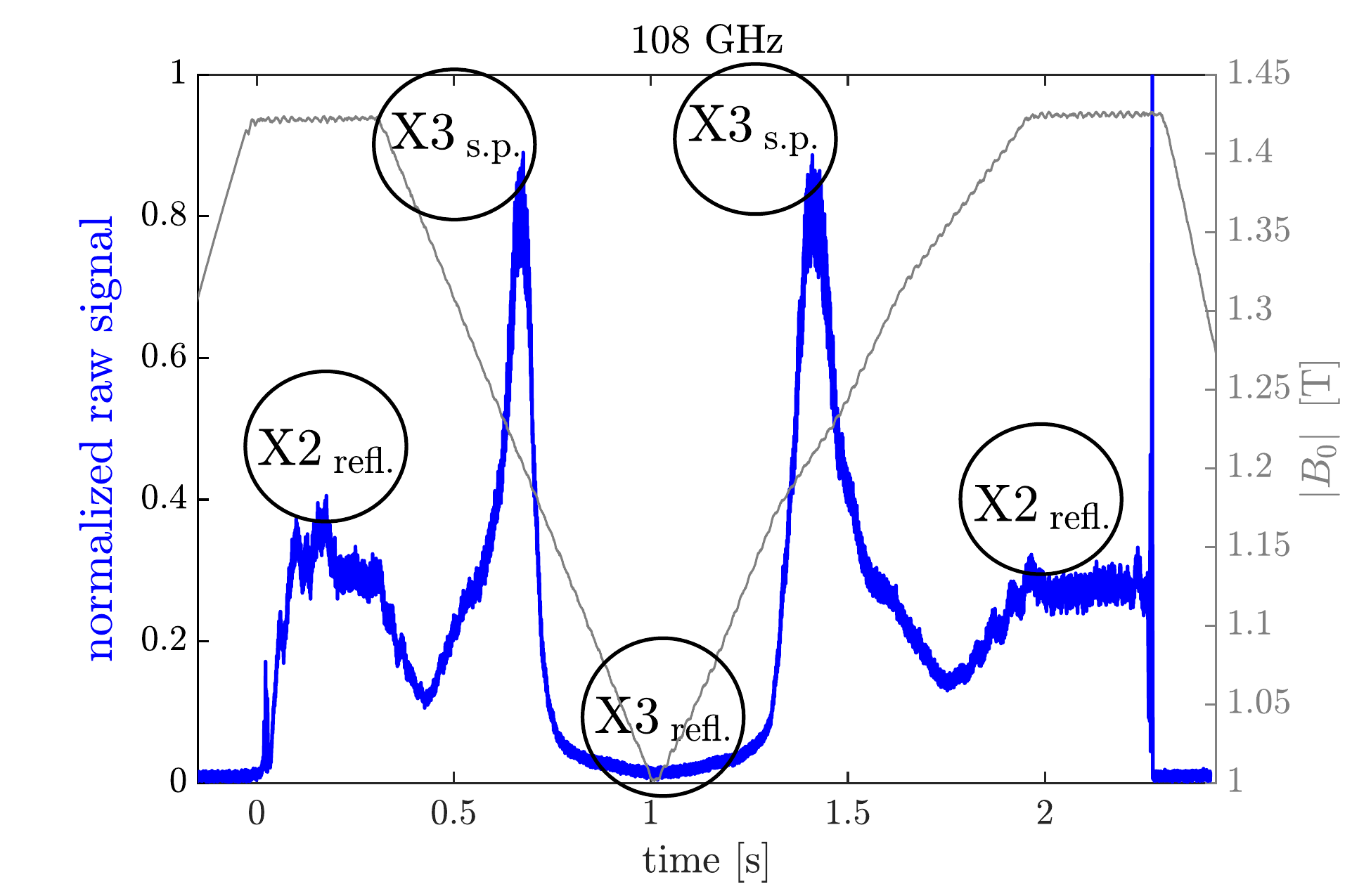} 
\caption[V-ECE signal measured at 108 GHz during TCV discharge $\# 67946$. The measurement shows radiation from single pass X3 and from multiple wall reflection.]{V-ECE signal measured at 108 GHz during TCV discharge $\# 67946$. The measurement shows single pass X3 (s.p) and radiation from multiple wall reflection (refl.).}
\label{fig:radio2_67946_zoom_chan3_an} 
\end{figure}

The calibration uses measurements from thermal plasmas on TCV. During the plasma discharge the magnetic field is varied (ramped down and up), to identify the spectral features of the radiation. The variation of the magnetic field strength also allows the identification of several ECE resonances, meaning that multiple channels can be calibrated within a single discharge. 
The measurement in Figure \ref{fig:radio2_67946_overview} shows X-mode intensity taken during TCV discharge $\#67946$, in the frequency range $104-114$ GHz. The magnetic field is ramped down from  $\sim 1.4$T to  $\sim 1$T and up again,  while the plasma current remains nearly constant. Peaks are observed, consecutively in each channel during the field ramps. The peaks are symmetric and separated by a period during  which the signal intensity is beneath the noise level of the radiometer. We observe approximately the same signal level for the peaks during the ramp down and the ramp up of the field.  The intensity at the magnetic field flat-top is higher for the lowest frequencies during the first  flat-top. In the second flat-top, at the end of the discharge, the lower signal level can be attributed to a drop in plasma optical depth due to reduced density.
Figure \ref{fig:radio2_67946_zoom_chan3_an} illustrates, for the frequency $108$ GHz the spectral features of the measured signal. The two peaks in the signal  are identified as single pass emission from the third harmonic. Radiation from multiple wall reflection,  which originated as an X$3$ emission,  is below  noise level  in between the two peaks. The radiation from multiple wall reflections, which originated as an X$2$ emission is observed outside the peaks.    
The diagnostic measures both reflected radiation and single pass radiation as the magnetic field is varied. When present, the reflected X$2$ radiation largely dominates the background contribution to the measurement. When the magnetic field is lower than the range in which the single pass X$3$ emissions is detected, the signal intensity drops to below the noise level confirming the absence of the background radiation originating as an X$3$ emission. We are certain that the reflected thermal X$2$ radiation cannot pollute the single pass X$3$  emission because the  X$2$ cold resonance location is out of the plasma when that of the X$3$ traverses the line of sight. These observations suggest that the calibration can rely on  modeling  of only  single pass X$3$ radiation measured by the radiometer. 
\begin{figure}[ht!] 
\centering 
\includegraphics[width=1.\columnwidth]{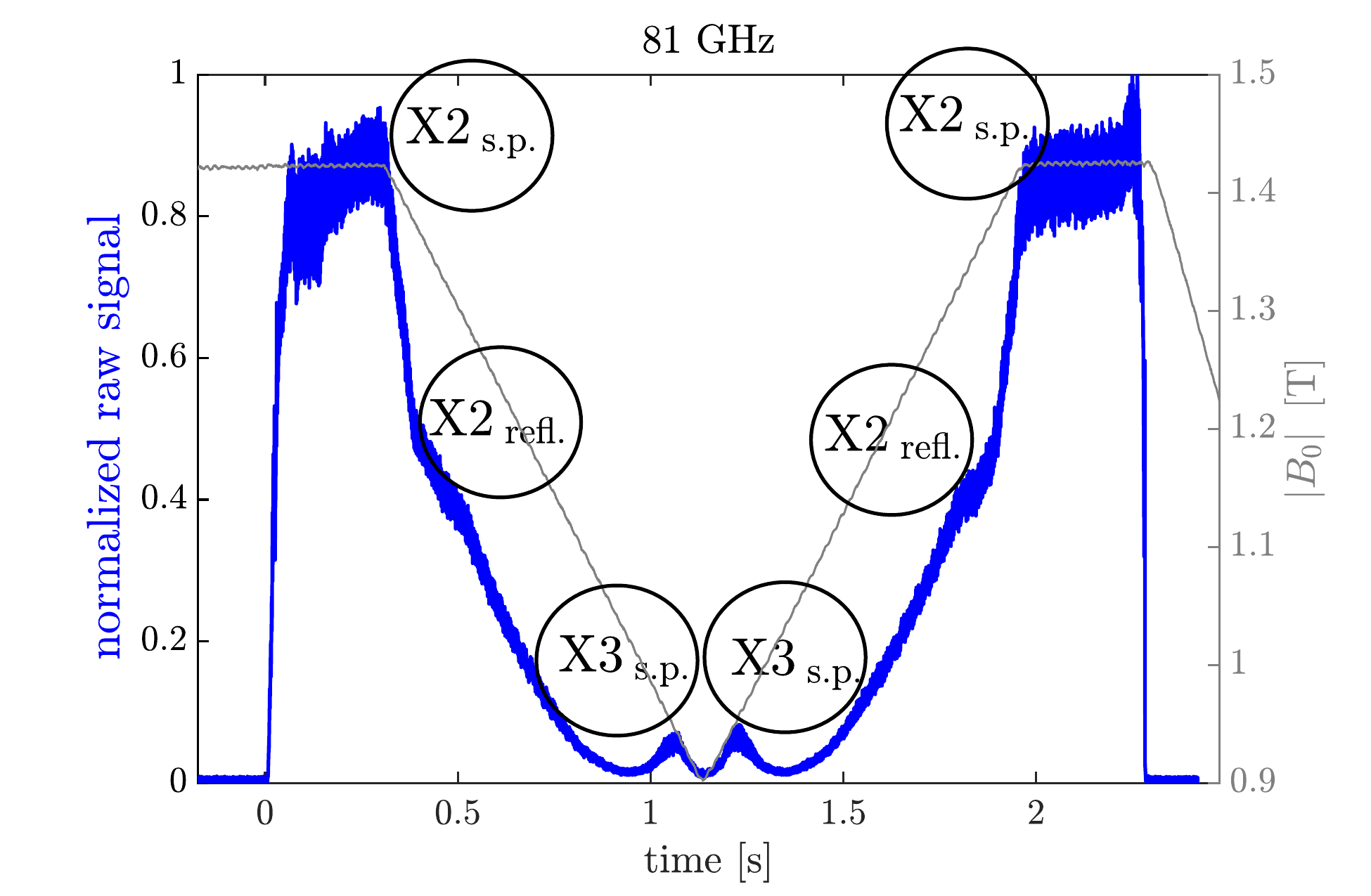} 
\caption[ V-ECE signal at lower frequency ($81$) during TCV discharge $\#75022$, showing detection of both X2 and X3 single pass radiation.]{V-ECE signal at lower frequency ($81$) during TCV discharge $\#75022$, showing detection of both X2 and X3 single pass radiation. }
\label{fig:radio51_75022_zoom_chan2_an} 
\end{figure}

We note that the lowest frequencies of the diagnostic can, in a single discharge, detect the single pass emission from both X$2$ and X$3$ $($Figure \ref{fig:radio51_75022_zoom_chan2_an}$)$. The single pass X$2$ radiation, which is optically thick and reaches the black-body level whose intensity depends solely on the electron temperature. For this reason, X$2$ single pass measurement can be used to verify modeling of X$3$ V-ECE radiation . 
\subsection{Analysis}
\label{analysis}
The modeling  aims at providing an estimation of the radiated thermal power in each frequency bandwidth.  The modeled power represents the power leaving the plasma to  ensure that all  components from the antenna to the detection system are included in the calibration. The modeled intensity from SPECE is the spectral intensity $I_{\omega}$ in Equation \ref{eq:93}. Integration over the bandwidth, effective area of the antenna and solid angle is performed to obtain the  power in each bandwidth. Since the antenna pattern is modeled in SPECE with a set of rays, the size of the ECE emitting layer in the plasma, relative to the antenna beam size was assessed by calculating the profile of the absorption coefficient with analytical expressions from \cite{Bornatici1982}.  
The modeling  is performed accounting for the fact that the X$3$ emission layer is narrower than the antenna beam width. An optimization process was undertaken to determine the optimal number of rays and other parameters of the model. A scan of the total number of rays suggested a minimum of 24 rays to achieve convergence of the mean spectral intensity. The number of rays used in this modeling is distributed in SPECE as 7 concentric annular sets, with 4 rays in each set.  Scans of the line of sight around the vertical direction showed that an inclination of approximately 3$^{\circ}$ toward the high field side would reproduce best the measurement. The beam size  used in the model is determined  by comparing the width of the X3 and X2 resonances as they pass through the beam optics. A beam size of 3 cm is found to  match the width of the resonances, consistent with beam size measurements in \cite{Biwole2023_vece}. For the calculation of the power in the bandwidth, the synthetic diagnostic computes and sums spectral intensities at discrete values of frequencies in each bandwidth. A scan in the frequency step $\delta_f$ led to a bandwidth discretisation with a minimum of $15$ frequencies $($$\delta_f \sim 53$ MHz$)$ per frequency band. 

\subsubsection{Calculation of the peak power in a frequency band\\}
\begin{figure}[ht!] 
\centering 
\includegraphics[width=1.\columnwidth]{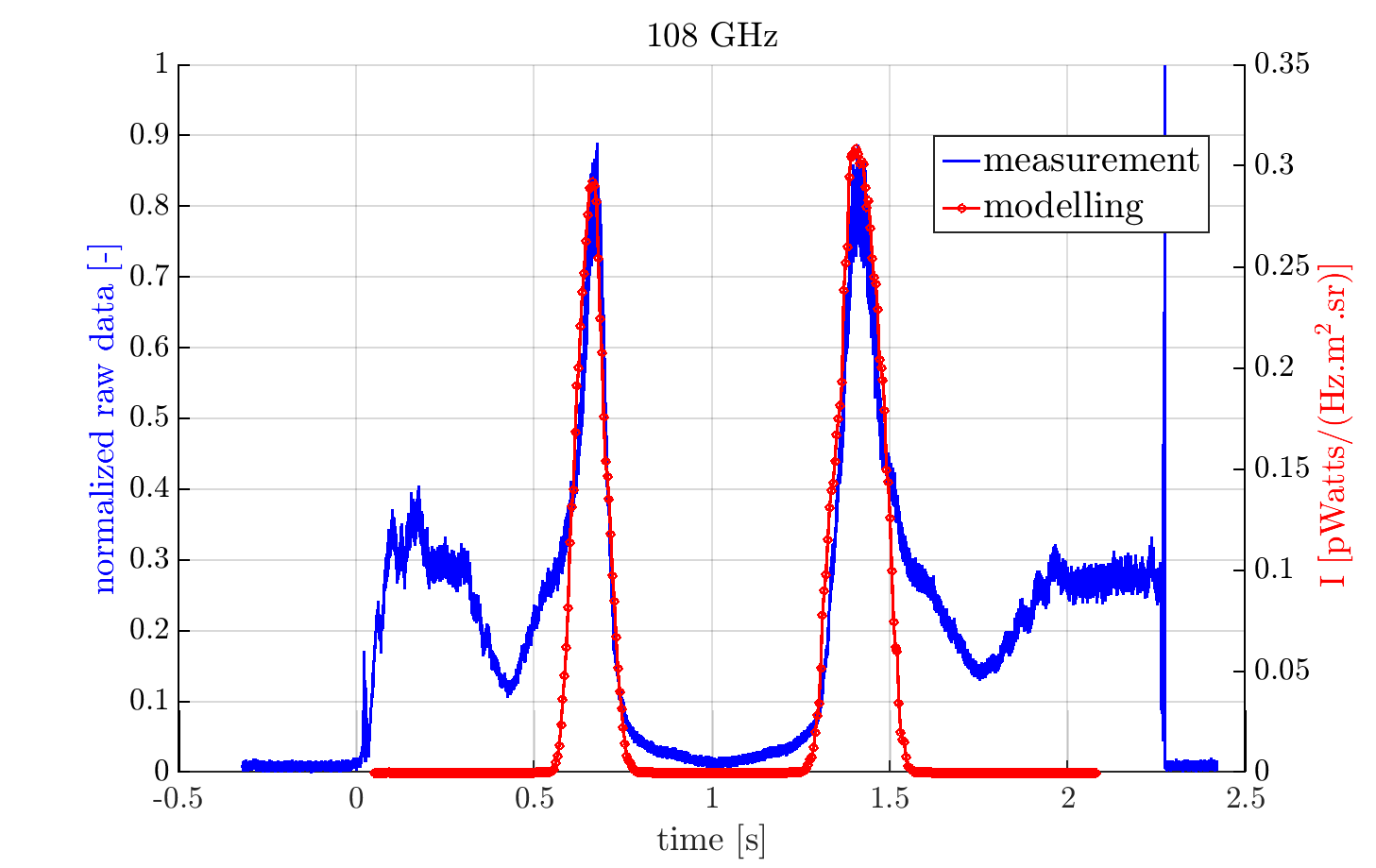}
\caption[Modeling of the spectral intensity for the shot $\#67946$ at frequency $108$ GHz ]{modeling  of the spectral intensity for the shot $\#67946$ at frequency $108$ GHz.}
\label{fig:radio1_67946_chan3} 
\end{figure}

\begin{figure}[ht!] 
\centering 
\includegraphics[width=.83\columnwidth]{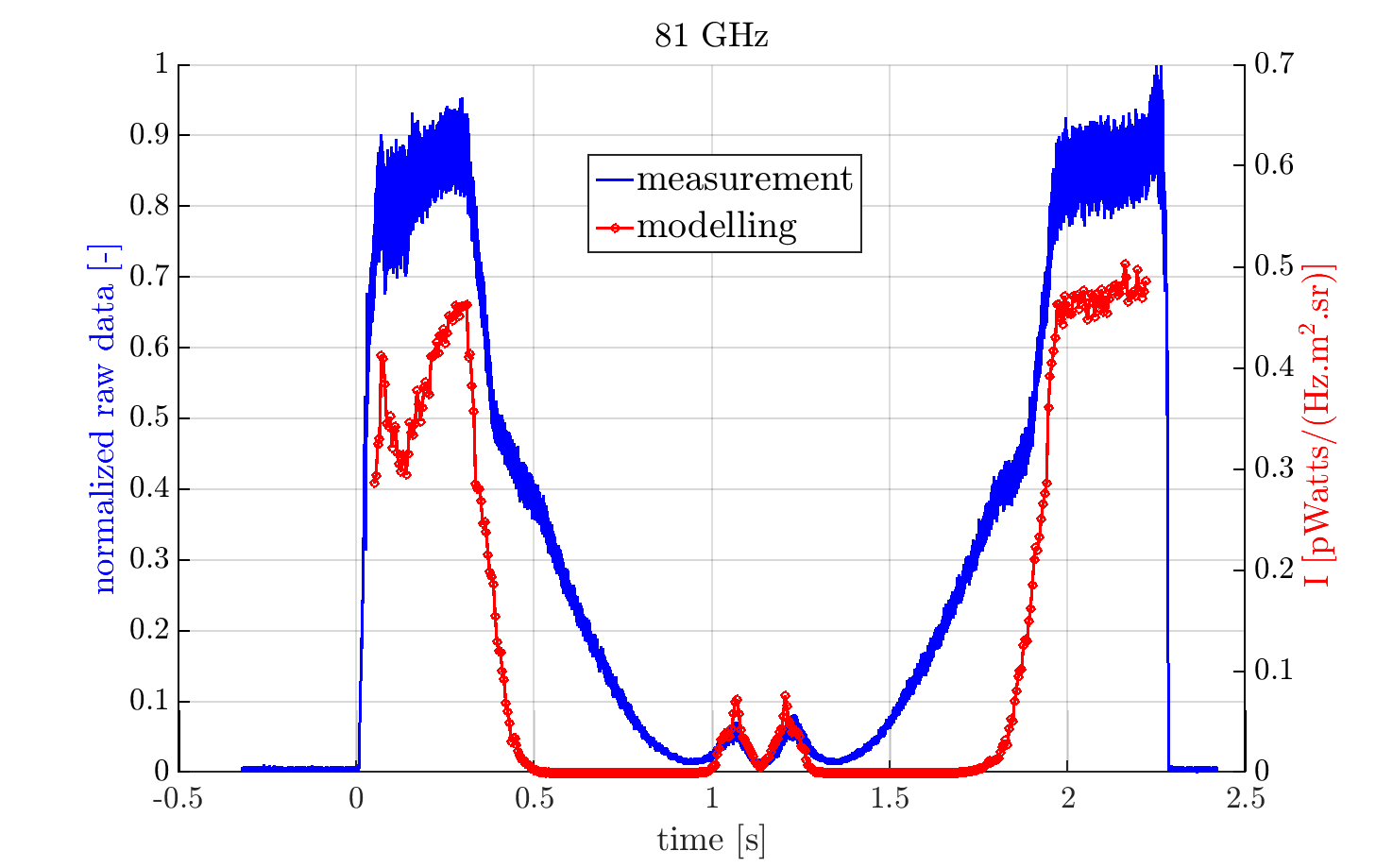} 
\caption[Modeling  of the spectral intensity for the shot $\#75022$ at frequency $81$ GHz ]{Modeling  of the spectral intensity for the shot $\#75022$ at frequency $81$ GHz.}
\label{fig:radio51_75022_chan2} 
\end{figure}

 The X- mode spectral intensities calculated for frequencies 108 GHz during TCV discharge $\#67946$ and 81 GHz during $\#67946$ are shown in Figure \ref{fig:radio1_67946_chan3}  and Figure \ref{fig:radio51_75022_chan2}. The results show good qualitative agreement between the modeling  of the single pass radiation and the measurement. The modeling  well predicts the temporal extent of the two peaks in $\#67946$. The modeling, as expected does not calculate the background radiation from multiple wall reflections.
The trends of the observed X$2$ and X$3$ single pass emissions are qualitatively reproduced  by modeling. The model captures the flat trend of the X$2$ single pass radiation towards the end of the discharge. The model also exhibits a drop in intensity at the beginning of the discharge consistent with the measured signal. 

\begin{figure}[ht!] 
\centering 
\includegraphics[width=.9\columnwidth]{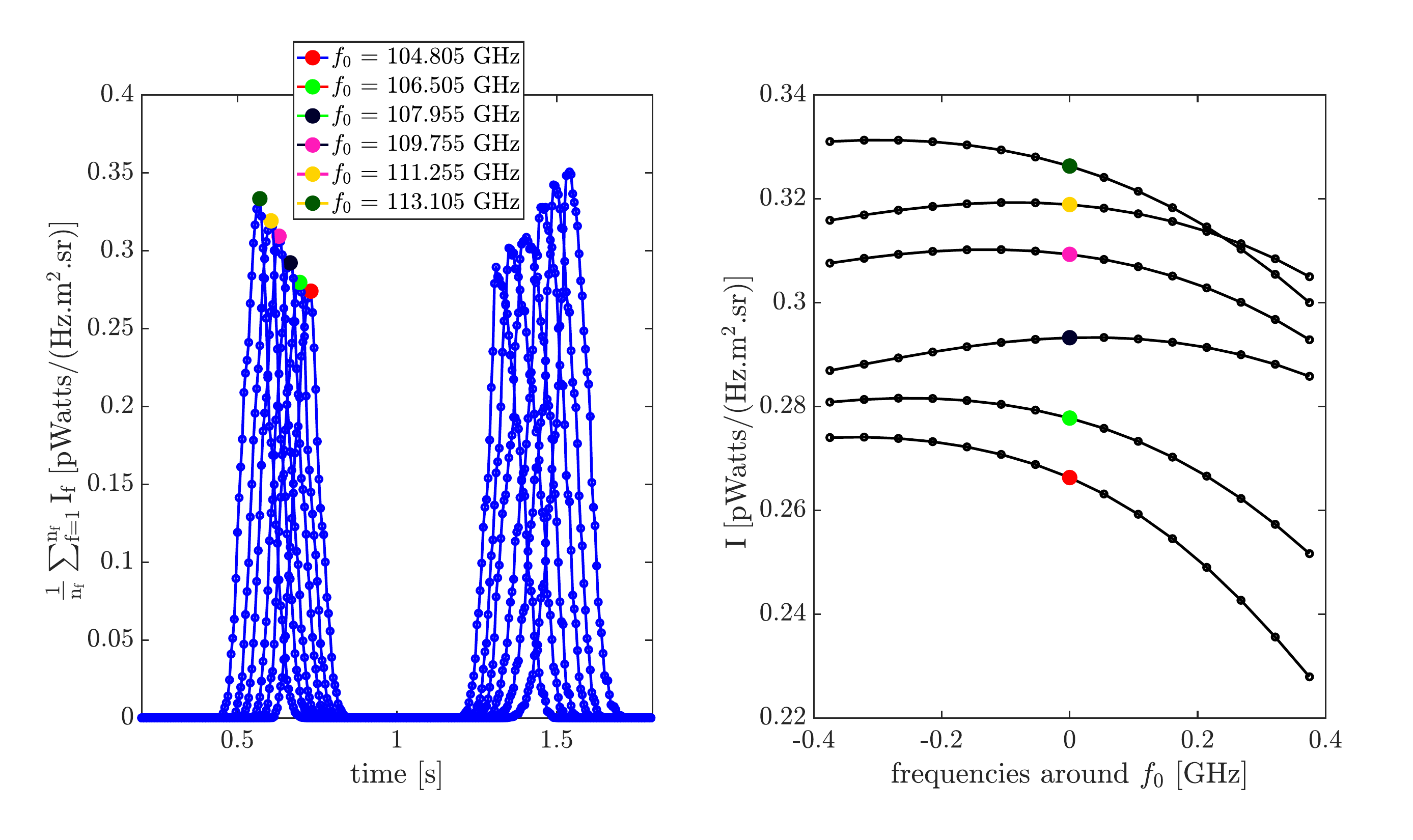} 
\caption[Modeled intensity in radiometer $104-114$ GHz during the shot $\#67946$ ]{Modeled intensity in radiometer $104-114$ GHz during the shot $\#67946$. }
\label{fig:mean} 
\end{figure}

The calibration factors are obtained by matching the modeled peak power with the measured voltage. The spectral intensities calculated for  a few frequencies during  discharge $\#67946$  are  shown in Figure \ref{fig:mean} with their profile in each bandwidth. The  parabolic shape of the intensity profile in the bandwidth justifies the need for  optimization of the frequency step. The bandwidth power $P_{\mathrm{BW}}$, the total power in each frequency channel, is calculated as:
 \begin{equation}
     P_{\mathrm{BW}} = 2 \pi \int df dA  d\Omega_{\mathrm{S}} I(f).
     \label{eq:bandpow}
 \end{equation}
From antenna theory, it can be shown that under certain approximation \cite{Shaw2012}, the product $dA  d\Omega_{\mathrm{S}}$ is only a function of the  radiation wavelength, $dA  d\Omega_{\mathrm{S}}\approx \lambda^2$. Equation \ref{eq:bandpow} thus read  
\begin{equation}
P_{\mathrm{BW}} \approx 2 \pi \int df \lambda^2 I(f) \approx 2 \pi c^{2} \int df \frac{I(f)}{f^{2}}.  
\label{eq:bandpow1}
\end{equation}

\subsection{Verification on the Calibration}
\label{valid}
\begin{figure}[ht!] 
\centering 
\includegraphics[scale=0.5]{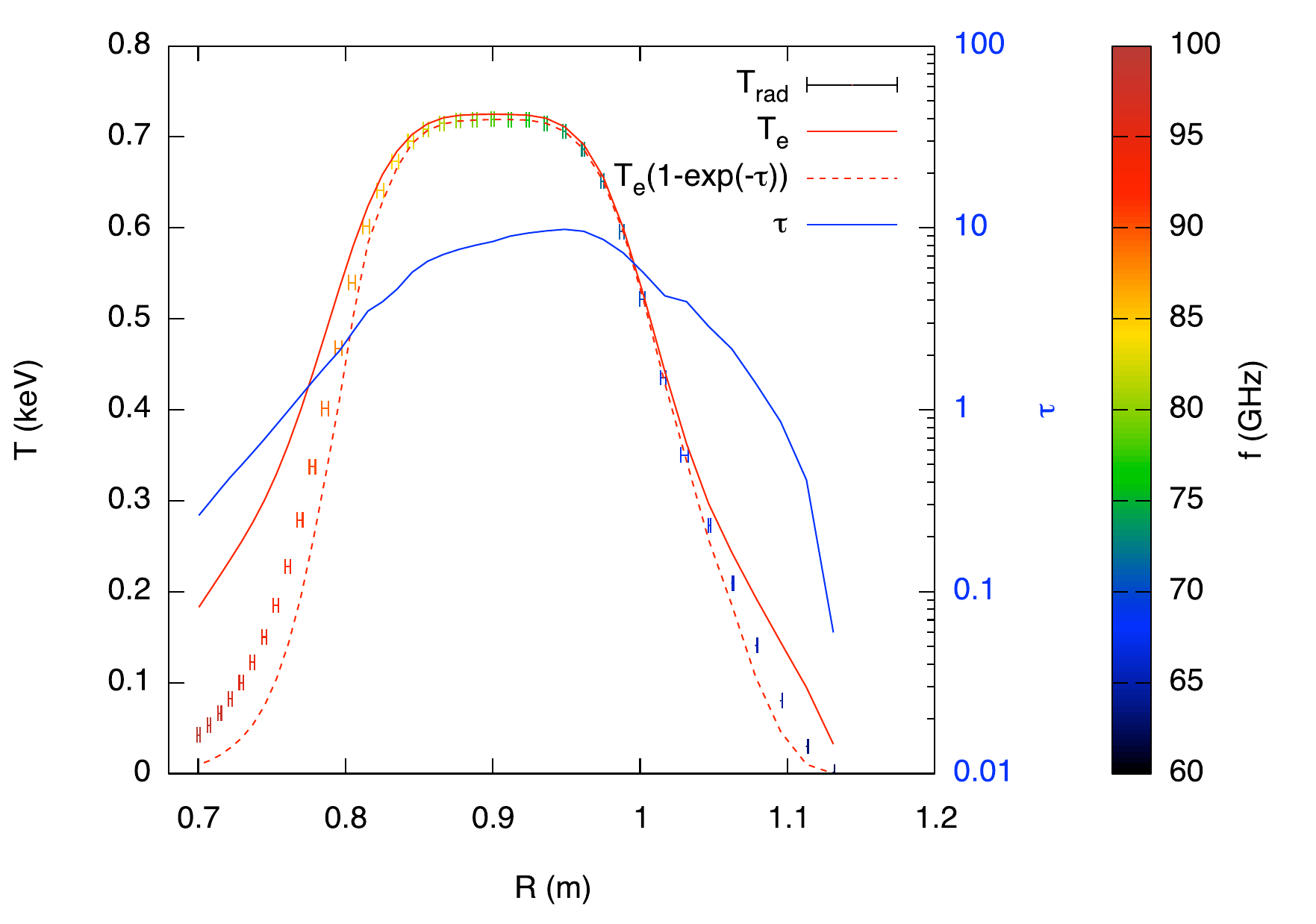} 
\caption[A result of the synthetic diagnostic applied to a horizontal line of sight:  the simulated  antenna temperature (Trad) follows the antenna temperature as it is  found analytically in a plasma with optical depth $\tau \sim 1$. In the Tokamak center, the simulated radiation temperature coincides with the local electron temperature as expected in optically thick thermal plasmas.]{A result of the synthetic diagnostic applied to a horizontal line of sight:  the simulated  antenna temperature (Trad) follows the antenna temperature as it is  found analytically in a plasma with optical depth $\tau \sim 1$. In the Tokamak center, the simulated radiation temperature coincides with the local electron temperature as expected in optically thick thermal plasmas.}
\label{fig:tevstrad} 
\end{figure}
SPECE  was used to estimate the bandwidth power. In an attempt to verify  SPECE  results on TCV, we first calculate the radial profile of the radiation temperature as measured by LFS ECE. The modeling  of the LFS ECE radiation intensity is consistent with expectations and is used as a sanity check of SPECE on TCV, see Figure \ref{fig:tevstrad}.

A verification of  V-ECE calibration with SPECE  is achieved with the plasma discharge $\#75022$ in which the diagnostic measures both X2 and X3 single pass radiation, Figures \ref{fig:radio51_75022_zoom_chan2_an} and \ref{fig:radio51_75022_chan2}. The calibration factors, resulting from modeling  of X3 intensity at $\sim 1.1$ s and $\sim 1.3$ s yield the intensity of the  black-body power measured around $2$ s (or around $\sim 0.3$ s ), $P_{\mathrm{BB}}  \sim 94$ nW. 
The mean electron temperature associated with this black-body radiation is estimated as $\overline{T_e} =  P_{\mathrm{BB}} / \Delta f_{\mathrm{BW}}$ $\sim 780 \; [\mathrm{ev}]$. Value which is close to the the central electron temperature  of  $\sim 1$ keV measured in that plasma. 
\subsection{Discussing the uncertainties}
\label{uncer}

\begin{figure}[ht!] 
\centering 
\includegraphics[scale=0.35]{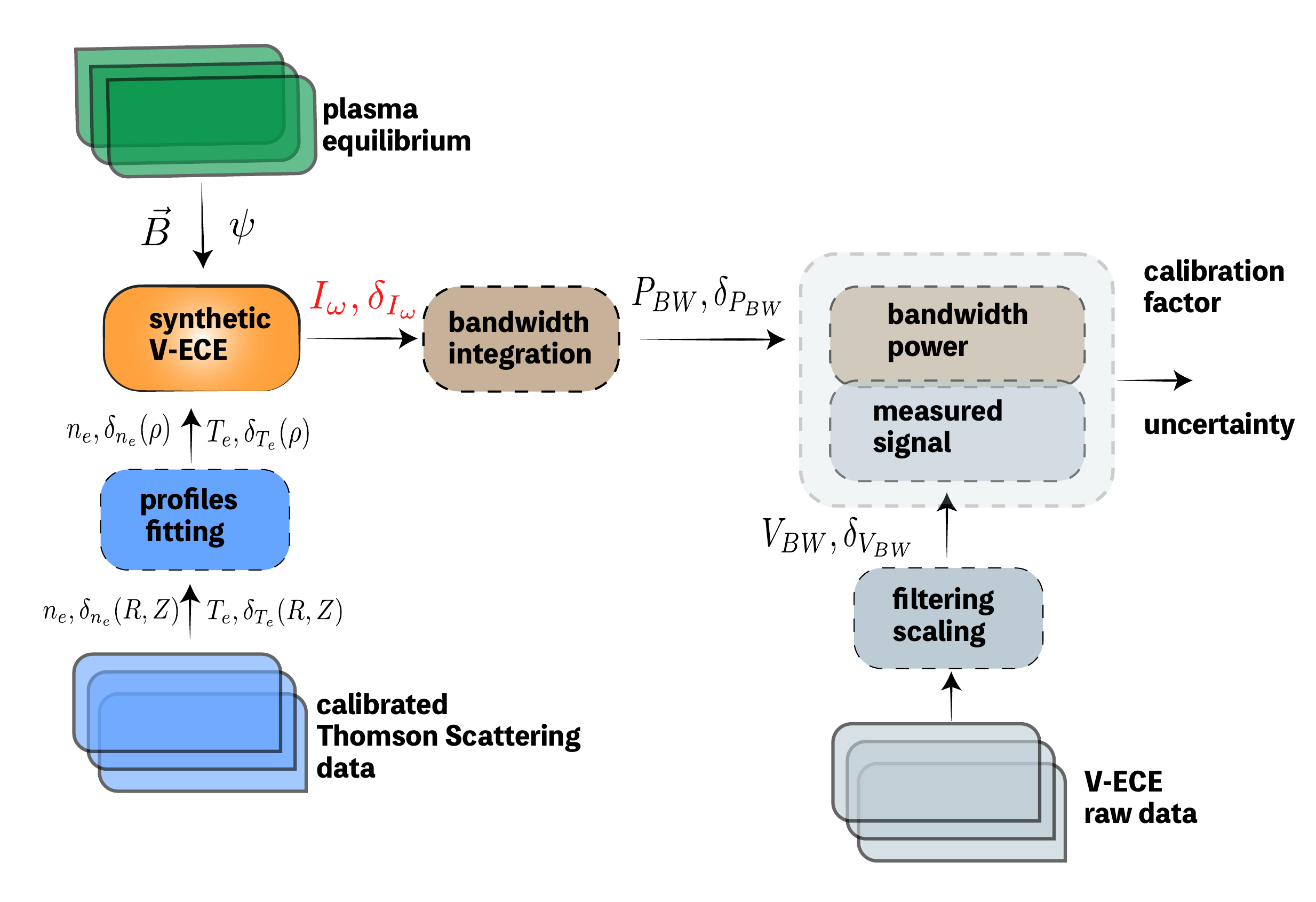} 
\caption[Block diagram illustrating the steps for the estimation of the uncertainty in the calibration procedure.]{Block diagram illustrating the steps for the estimation of the uncertainty in the calibration procedure.}
\label{fig:unce_steps} 
\end{figure}

The assessment of the uncertainty in the calibration factors is discussed for the case example of discharge $\#67946$. The steps of the assessment are summarized in the diagram of Figure \ref{fig:unce_steps}.  The steps are subdivided in two main branches. One branch for the uncertainty from the processing of the V-ECE raw signal. The other branch  for the uncertainties in the estimation of the bandwidth power using the synthetic diagnostic.

The uncertainty in the V-ECE signal primarily arises from the digital (Butterworth) filtering of the raw data. Butterworth filtering is applied to the data to eliminate some noise in the detected signal, typically introduced by adjacent electronic components.

For the case of interest, the relative uncertainty on the filtered value used for the calibration is $\delta_{V}/{V}  < 5 \%$. 
To assess the uncertainty in the modeled bandwidth power, the uncertainties in the parameters are propagated  from the raw measurement to the calculated bandwidth power, as shown in  in  Figure \ref{fig:unce_steps}. The assessment does not account for the potential uncertainty in the measurement of the magnetic field because  the diagnostic is calibrated with thermal peak measurement, when the  resonance is within the line of sight. The measured frequencies and the the line of sight  constrain the value of the magnetic field at the resonance. The uncertainties that are propagated are those from the measurement of the electron density, $\delta_{n_\mathrm{e}}$ and electron temperature, $\delta_{T_\mathrm{e}}$. On TCV,  The relative errors for both  density  and  temperature measurements with Thomson Scattering  are respectively below $5\%$ and $10\%$ in the plasma center. Both values increase significantly at the plasma edge, up to $25\%$ and $40\%$ respectively. The high uncertainties on the density and temperature profiles at the edge do not impact  significantly  the modeled intensity. That is because the optical depth of the plasma drops at the edge in such a way that the intensity depends mainly on the parameters in the  plasma center. 

\begin{figure}[ht!] 
\centering 
\includegraphics[width=0.8\columnwidth]{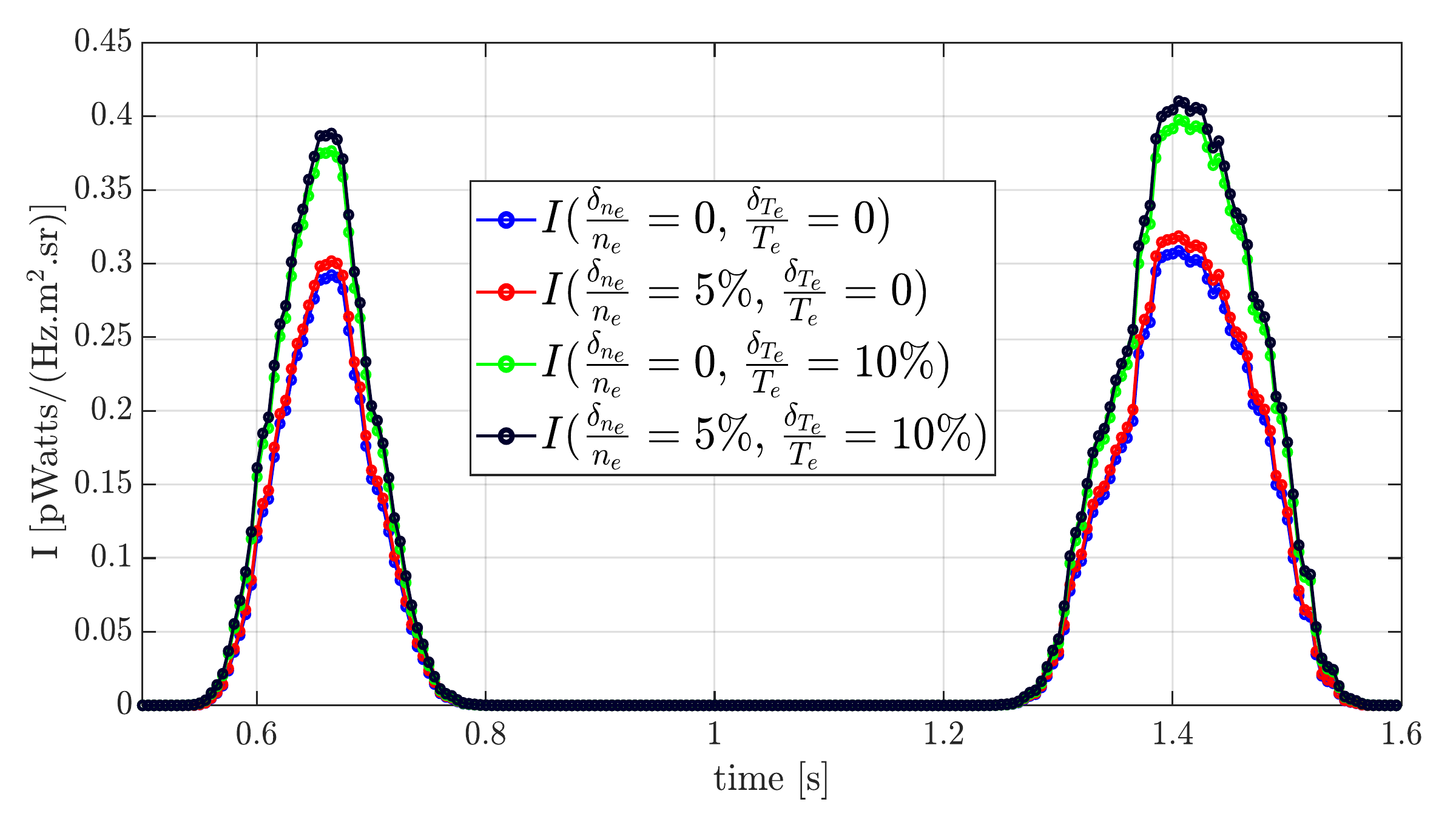} 
\caption[Spectral intensities calculated with plasma parameters affected by uncertainties.]{Spectral intensities calculated with plasma parameters affected by uncertainties.}
\label{fig:uncer_i_dn_dt} 
\end{figure}

The relative uncertainty, $\delta_I / I $, on the computed spectral intensity for a frequency of $104$ GHz is in the order of $\delta_I/{I} < 25 \%$.
Figure \ref{fig:uncer_i_dn_dt} shows the spectral intensities calculated with the synthetic diagnostic for different cases, with different uncertainties on the plasma profiles. The main contributor to the uncertainty on the spectral intensity is the uncertainty on the electron temperature. The uncertainty on the electron density plays a minor role.  This is  explained by the fact that the modeled radiation is single pass third harmonic intensity, which has a stronger dependence on the electron temperature. 
 
Assuming that the relative uncertainties in the spectral intensity depend only on the plasma parameters and not on the modeled frequencies, the uncertainties in the spectral intensity will propagate unaffected through the integral  of Equation \ref{eq:bandpow1}, giving an uncertainty on the modeled bandwidth power of $\delta_{P}/{P}  < 25 \%$. This sums up to an uncertainty on the calibration factor of the order of $\delta_{F}/{F}  < 30 \%$, accounting for the uncertainties in the modeled power and the uncertainties from the raw V-ECE signal. We note that this overall uncertainty on the calibration factor is essentially due to the systematic uncertainty on the raw electron temperature measured by the Thomson Scattering diagnostic.

\section{Fast electron measurement}
\label{meas4b}
\subsection{ECCD measurement}
\label{meas4b_meas}
\begin{figure}[ht!] 
\centering 
\includegraphics[width=1.\linewidth]{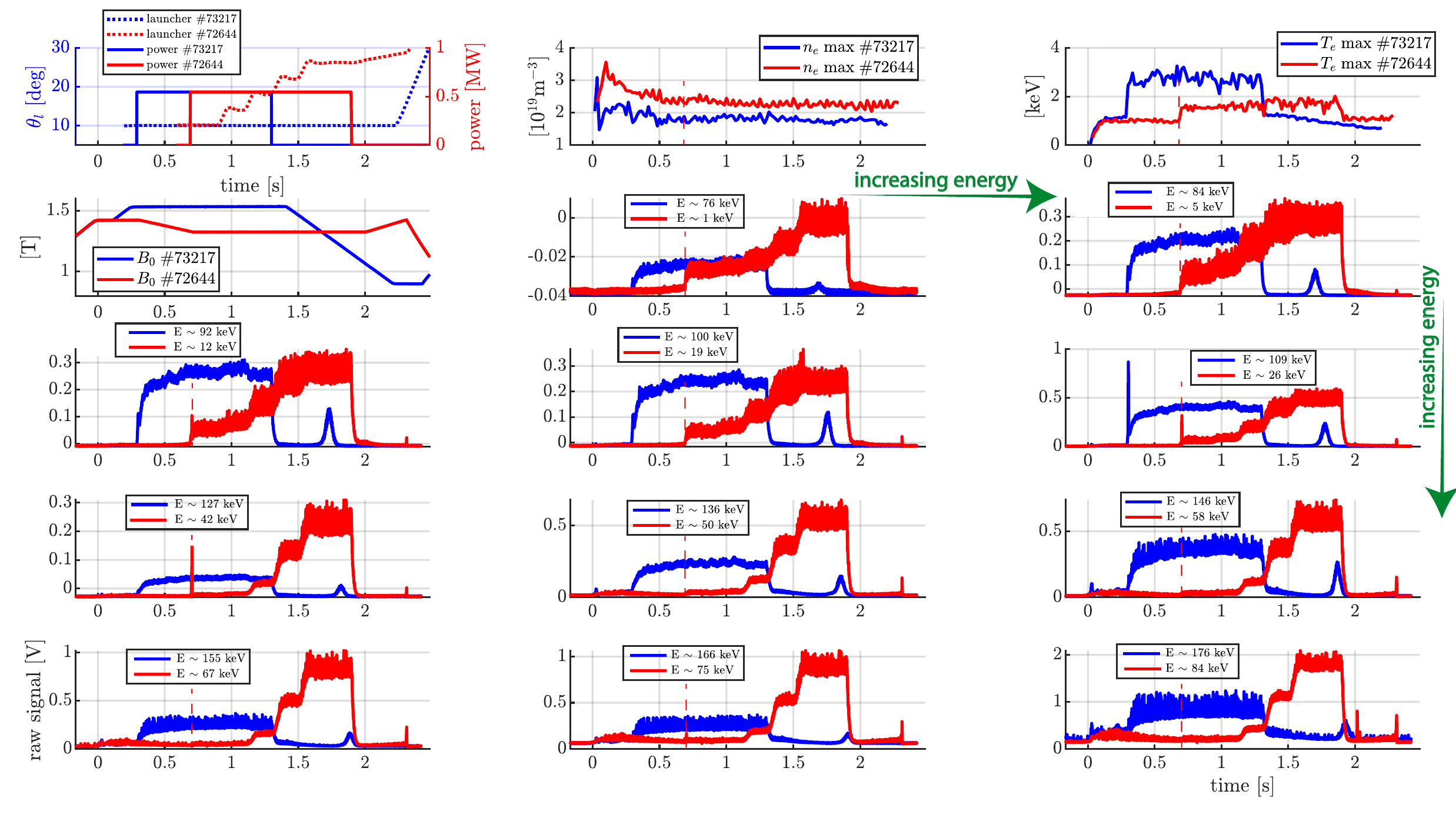} 
\caption[Raw measurements during the discharges $\#73217$ and $\#72644$ in the frequency range $96-114$ GHz]{Raw measurements of X-mode intensity from discharges $\#73217$ and $\#72644$ in the frequency range of $96-114$ GHz. The signals are plotted in decreasing frequency, from upper left to lower right. The energy measured at each frequency is calculated assuming a measurement of downshifted radiation from the third harmonic in both discharges. The subplot in the second row, second column, corresponds to the highest frequency and lowest energy (114 GHz, $\sim$76 keV in $\#73217$ and $\sim$1 keV in $\#72644$), while the subplot at the bottom right corresponds to the lowest frequency and highest energy (96 GHz, $\sim$176 keV in $\#73217$ and $\sim$84 keV in $\#72644$). The calculated energy differs between the two discharges due to the different values of  magnetic field strength during the heating phase. The vertical dashed line represents the time when ECH heating starts in discharge $\#72644$. The signal jump occurs farther from the vertical line at the highest energies.}
\label{fig:raw_2shots} 
\end{figure}
\begin{figure}[ht!] 
\centering 
\includegraphics[width=1.\linewidth]{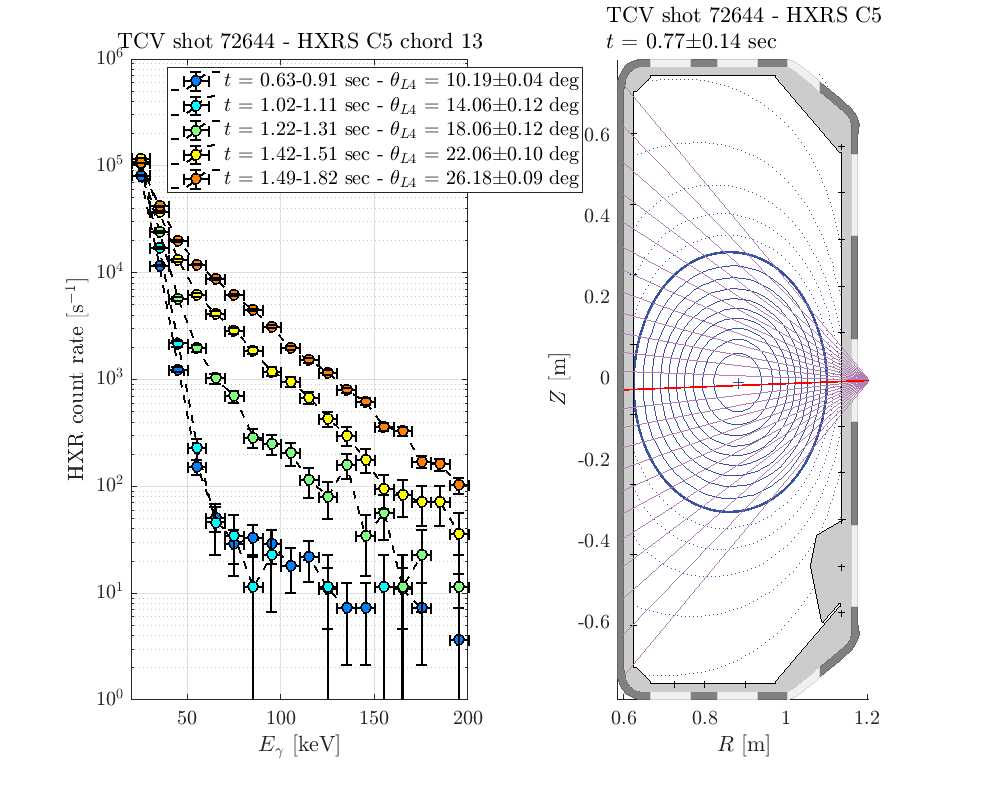} 
\caption[Hard-X-Ray data for TCV discharge $\#72644$, showing the effect of the change in the launcher poloidal angle]{Hard-X-Ray data for TCV discharge $\#72644$, showing the effect of the change in the launcher poloidal angle. Higher count rates are measured at higher energies for increasing  launcher poloidal angles. }
\label{fig:72644_hxrs} 
\end{figure}
\begin{figure}[ht!] 
\centering 
\includegraphics[scale=0.3]{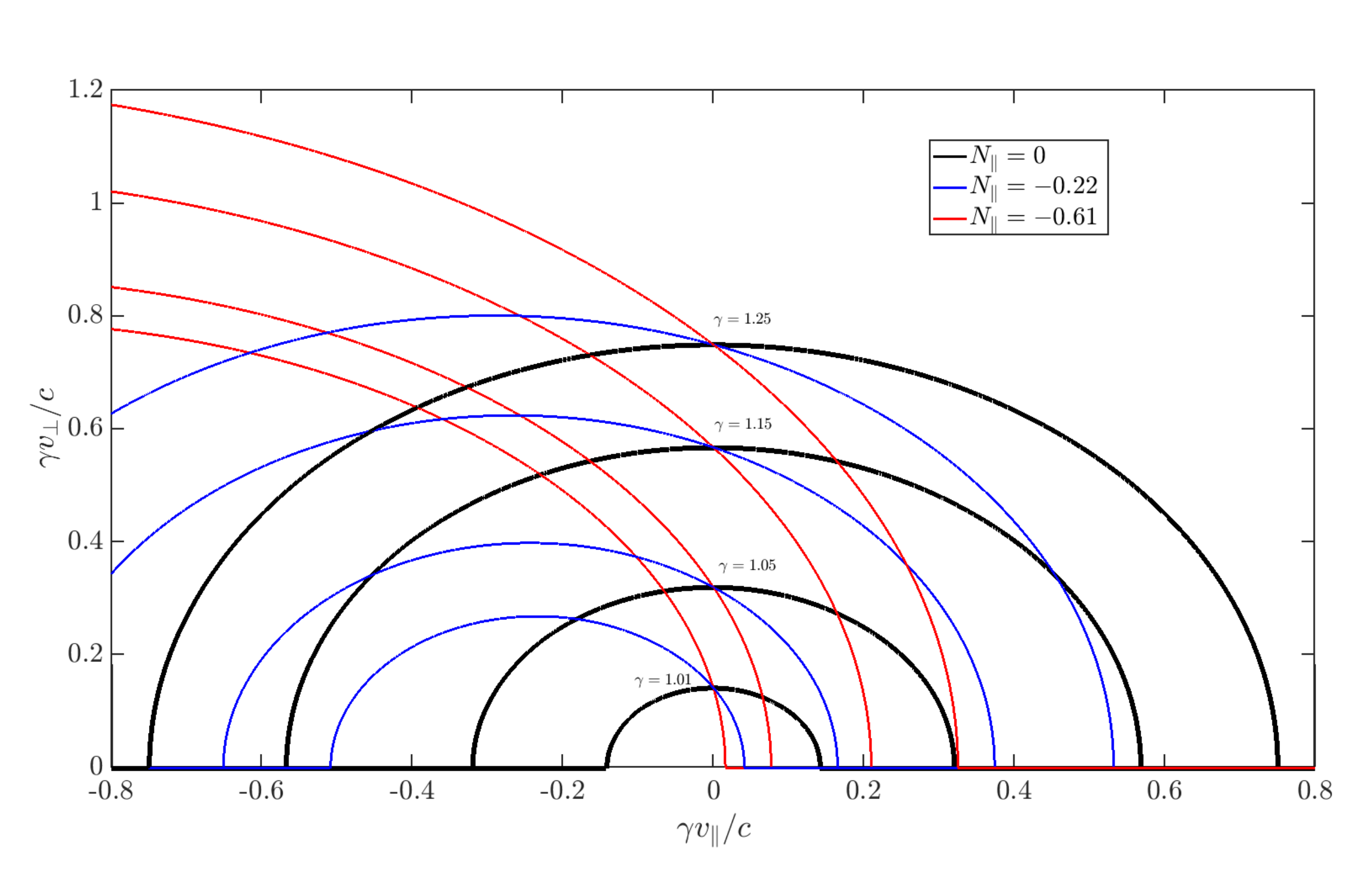} 
\caption[Electron Cyclotron Resonance curves for different values of parallel refractive index and energies.]{Electron Cyclotron Resonance curves for different values of parallel refractive index and energies.}
\label{fig:upar_uperp} 
\end{figure}
\begin{figure}[ht!] 
\centering 
\includegraphics[scale = 0.3]{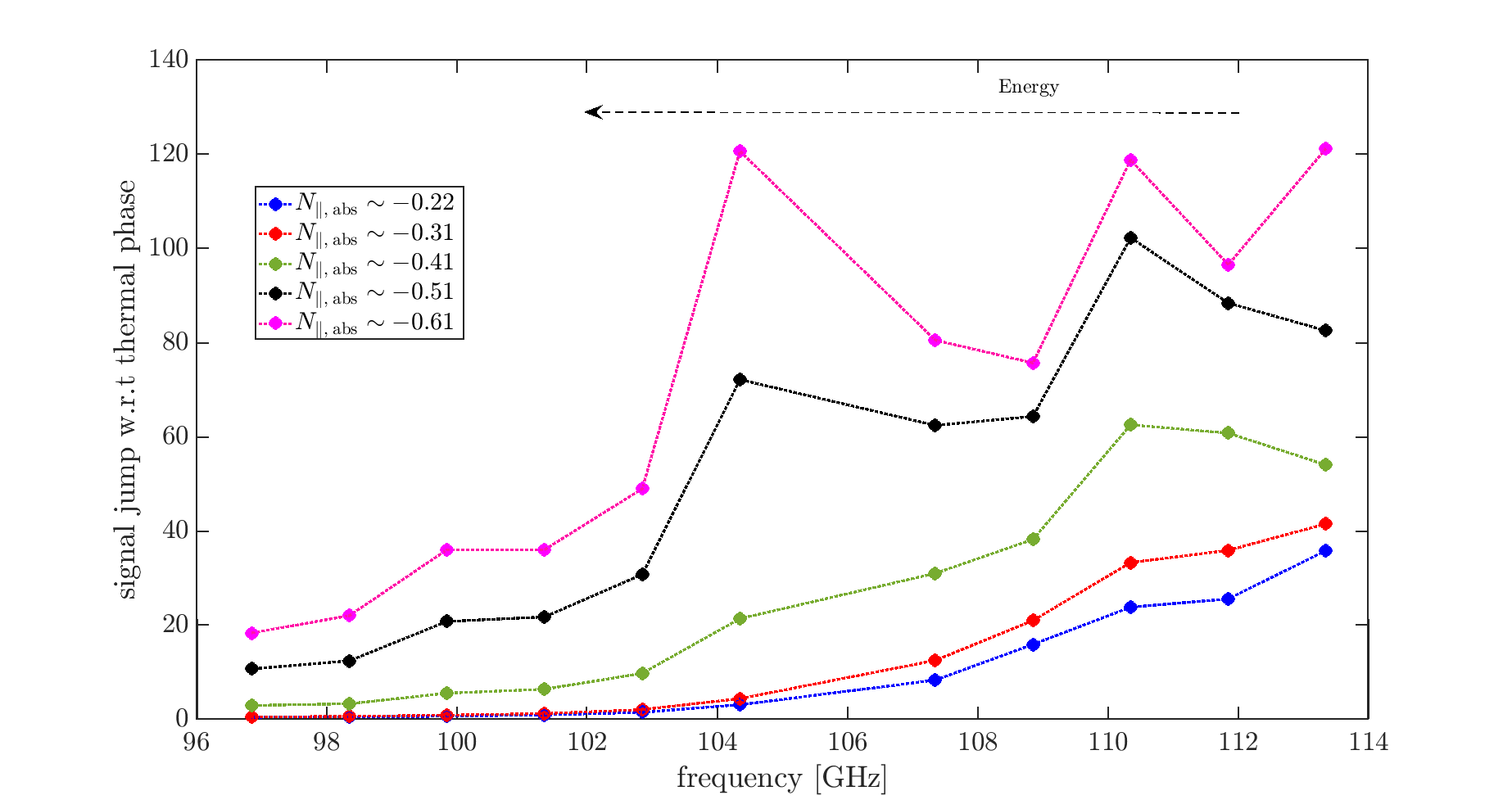} 
\caption[X-mode intensity as a function of energy at different values of parallel refractive index]{X-mode intensity as a function of energy at different values of parallel refractive index.}
\label{fig:new_sauter} 
\end{figure}
\begin{figure}[ht!]
\centering    
\subfloat[]
{\label{fig:plot_a}%
\includegraphics[width=0.485\columnwidth]{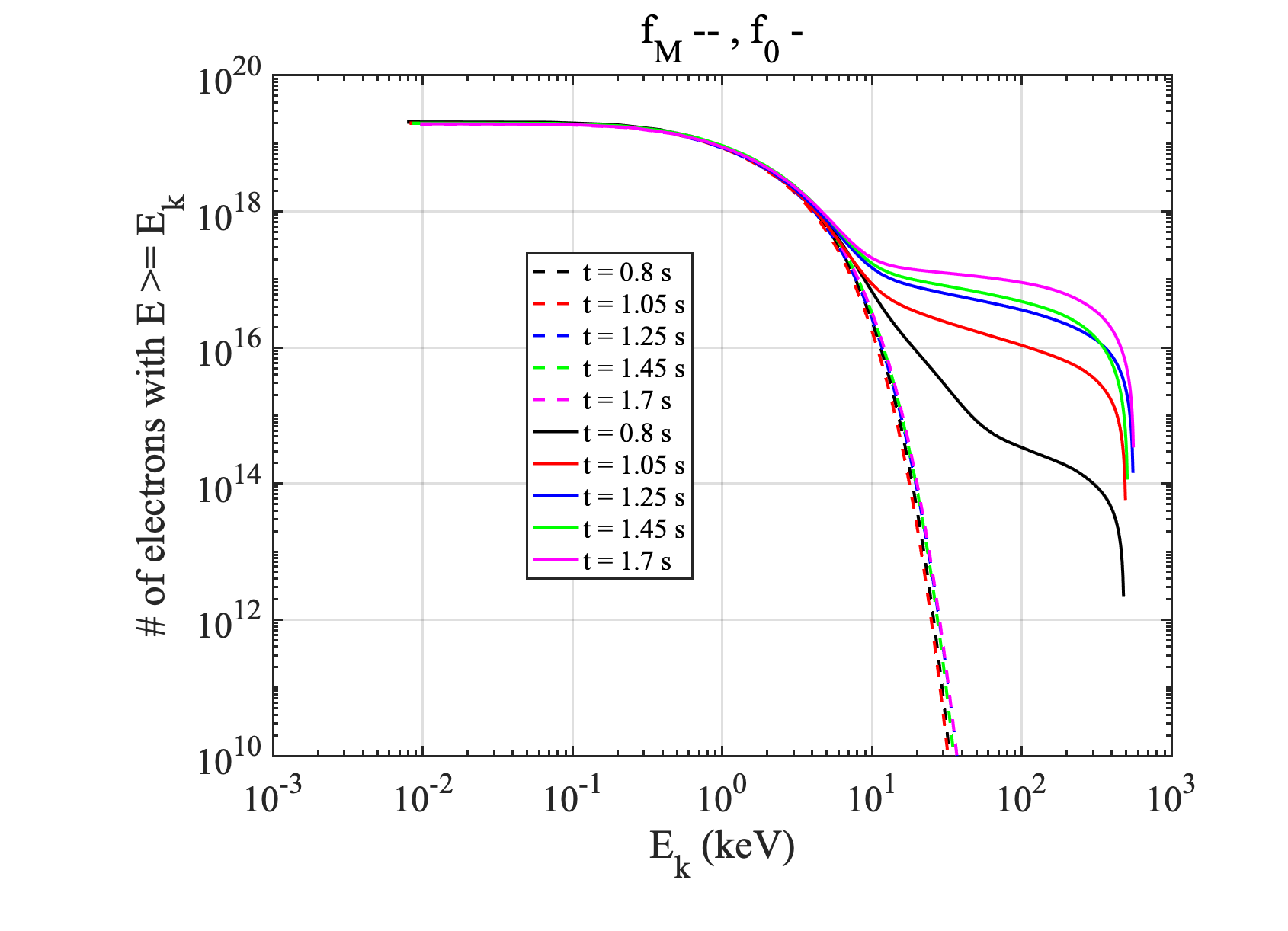}} \quad
\subfloat[]
{\label{fig:plot_c}%
\includegraphics[width=0.485\columnwidth]{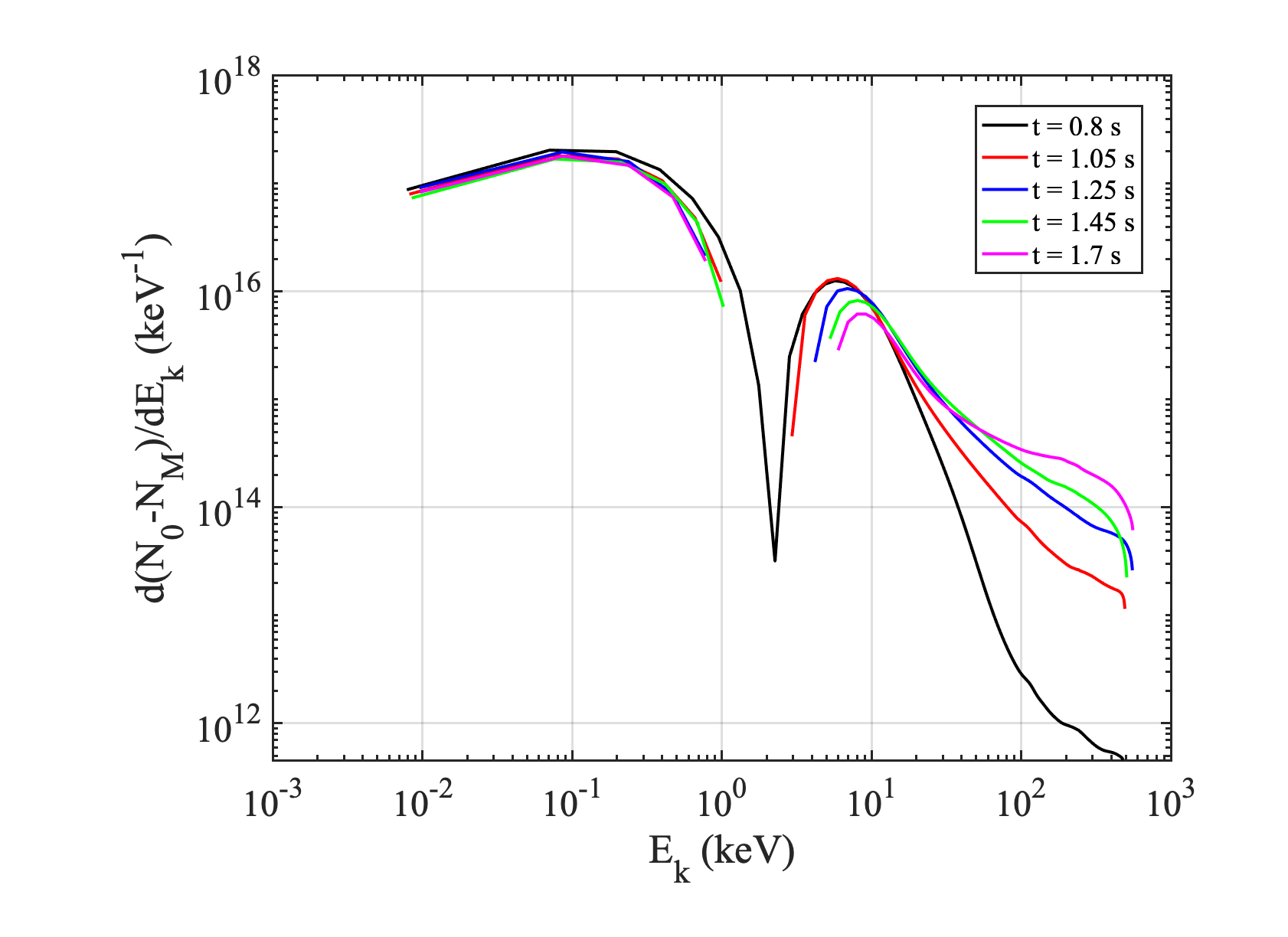}} 
\caption[]{\ref{fig:plot_a} Calculation with the code LUKE of the total number of electrons above a certain energy. The dashed lines are relative to the Maxwellian distribution. The solid lines are the computed distribution for the discharge. Derivative of number of electrons with respect to energy with respect to the Maxwellian distribution \ref{fig:plot_c} }
\label{fig:plot_luke}
\end{figure}

Calibrated V-ECE measurements during ECCD discharges $\#73217$ and $\#72644$ on TCV have been  presented in Reference \cite{Biwole2023_vece}. Raw measurements during those discharges are shown in Figure \ref{fig:raw_2shots}. Both discharges have similar values of ECH power. While the launcher toroidal angle is kept constant, the poloidal angle is varied in $5$  steps during the heating phase between $0.7$ s and $1.9$ s in $\#72644$. In this configuration, as the EC beam propagates along the plasma equatorial mid-plane, the launcher angle controls the parallel component of the wave vector with respect to the toroidal magnetic field in the plasma. When this parallel wave-vector component is zero ($\phi_{\mathrm{L}} = 0$), the EC wave purely heats the plasma by increasing the perpendicular energy of the resonant electrons around a given parallel velocity. Increasing the parallel wave vector (by increasing  $\phi_{\mathrm{L}}$ )  results in a Doppler shift of the EC resonance, pulling a suprathermal electron tail out of the bulk at higher velocities than the corresponding heating case \cite{prater_2004}. The variation of the poloidal angle of the launcher from $\sim 10 ^{\circ}$ to $\sim 26 ^{\circ}$ in each step allows the observation of a stair-shaped  X-mode radiation intensity. When no heating power is applied in either  plasma discharge, the measured radiation has similar intensity levels in both discharges, corresponding to the thermal background radiation originating from X2 emission.   As soon as the heating is turned on in the discharge $\#73217$ at $\sim 0.3$ s, all the channels immediately peak to reach a nearly stationary intensity level. We note how the intensity level during the discharge  $\# 73217$ is generally higher than the intensity level in the discharge  $\# 72644$ when the launcher configurations are similar. This  is due to the more efficient current drive achieved in $\# 73217$, due in part to the lower density  and higher electron temperature as the current drive efficiency is proportional, in general, to the ratio $T_e / n_e$ \cite{LinLiu2003}.  When the heating is turned on during the discharge  $\# 72644$,  at $\sim 0.7$ s, only the low energies $($$\sim$ $5-26 $ keV$)$ exhibit a sharp jump in the measured intensity. The intensity at the higher energies ($\sim$ $67-75 $keV) shows a similar jump, with a delay of more than $\sim 500$ ms, when the launcher angle has been  moved to its third stage. From that moment, the dynamics is inverted and the higher energies now have the sharpest jump in intensity compared to the lower energies. Interestingly, the measurement at the middle energies,  in our range, will then reach the highest value of intensity when the launcher angle reaches its last stage. In measurements  with Hard X-Ray tomographic Spectroscopy system  (HXRS) \cite{gnesin_hxrs} during the discharge $\# 72644$, the count rates  shown in Figure \ref{fig:72644_hxrs} allow similar observations on the effects of the launcher angle sweep. In TCV, hard x-rays are emitted by suprathermal electron bremsstrahlung and are measured between 20 and 200 keV with an 8 keV resolution. In Figure \ref{fig:72644_hxrs}, the sampling interval is set to 10 keV, which aligns with the resolution, making finer sampling unnecessary
Simulations of the ECH wave interaction with the plasma,  for the discharge  $\#72644$  is achieved using the the ray tracing code C3PO, solving the linear wave power absorption \cite{Peysson_2012}. 
The refractive indices and the  angle between the wave vector and the magnetic field, calculated at the wave absorption location  for the various launcher angles are shown in Table \ref{tab:angle_npar}. The increase in the launcher poloidal angle allows to achieve higher values $($in absolute terms$)$ of  the parallel refractive index. Higher parallel refractive indices generally allow the intersection between the EC resonant curves ($v_{\perp}$ vs $v_{\parallel}$)  and the $v_{\parallel}$ axis  at higher  values of $v_{\parallel}$ \cite{Poli2019}, see Figure \ref{fig:upar_uperp}. The physical mechanisms of the plasma-wave interaction mechanisms include, in general,  momentum diffusion and pitch-angle scattering, which make the ECCD resonant interaction region complex to determine. For the purposes of this experiment it is observed   that higher values $\theta_L$ and thus higher absolute values of $ N_{\parallel}$ induce more energetic electrons in the plasma, explaining the different jumps in the measured intensities.\\ 
\begin{table}[ht!]
    \centering
    \begin{tabular}{|c|c|c|c|c|c|}
    \hline
    Launcher $\phi_l$ [$^{\circ}$] & -90 & -90 & -90 & -90 & -90 \\ \hline
    Launcher $\theta_l$ [$^{\circ}$] & 10 & 14 & 18 & 22 & 26 \\ \hline
    $N_{\parallel}$ & -0.22 & -0.31 & -0.41 & -0.51 & -0.61 \\ \hline
    $\angle(\mathbf{k}, \mathbf{B})$ [$^{\circ}$] & -16.7 & -24.1 & -33.05 & -43.3 & -55.1 \\\hline
    \end{tabular}
    \caption{Summary of ECH parameters during TCV discharge $\#72644$}
    \label{tab:angle_npar}
\end{table}
\\Figure \ref{fig:new_sauter} show the jumps in  intensity, relative to the thermal phase, measured at each frequency in the different stages of discharge $\#72644$. The jump is obtained by dividing the intensity in each heated stage by the intensity in the thermal phase. It is seen how at relatively low parallel refractive index ($\left|N_{\parallel}\right| = 0.22$ ), the highest frequencies $($lowest energies$)$ jump is the most important. As $\left| N_{\parallel} \right|$  increases, an intermediate frequency (intermediate energy as well), near $104$ GHz shows the sharpest jumps. We note that V-ECE X-mode intensity in ideal conditions depends on the density of the non-thermal electrons in a given energy band. That is, the highest jump observed around $104$ GHz, with increasing $\left|  N_{\parallel} \right|$ suggests an increase in the number density of non-thermal electrons at the corresponding energy.  This does not necessarily mean that the ECH wave is damped on electrons whose V-ECE energies are measured at $104$ GHz when the refractive index increases. The increase of the number density at the given energy, which explains the sharp jump, is a consequence of the overall dynamics in the electron energy space, originally caused by the ECH wave.
First-principle calculation using the code LUKE is used compute the electron energy distribution at the different stages of discharge $\#72644$. The first calculation is shown in Figure \ref{fig:plot_a}. The total number of electrons above a certain energy is plotted  for a Maxwellian distribution and for the computed distribution. At a given energy, say 15 keV, which is above the electron thermal energy, the total number of electrons above that energy increases over time intervals (with increasing $\left| N_{\parallel} \right|$).  This results confirms that low-energy electrons are pushed into the fast-electrons tail with ECCD and the number of transported electrons increases with parallel refractive index. 
Figure \ref{fig:plot_c} shows the derivative of the number of electrons in the modeled distribution deduced from the density in the Maxwellian distribution. 
The discontinuity in the spectra is due to the logarithmic scale and to the absolute value in the plotted quantity. It represents the regions where the density in the Maxwellian distribution changes from higher to lower than that of the modeled distribution, which is  the region in the phase space where the electrons are being pumped out by the ECH wave.   
\subsection{Runaway electron measurement}
\begin{figure}[ht!] 
\centering 
\includegraphics[scale = 0.5]{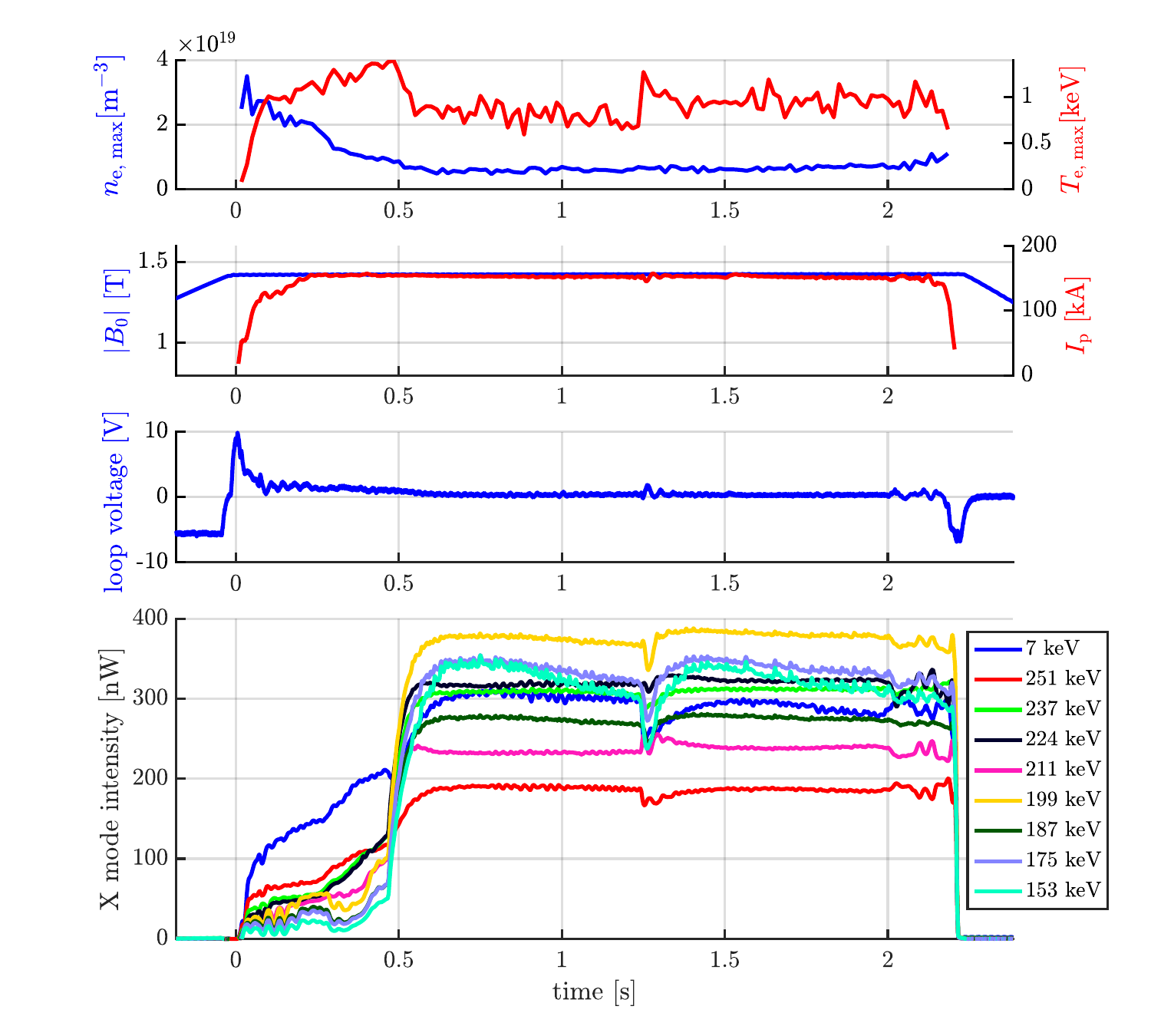} 
\caption[Time histories of runaway electron discharge $\#73002$ featuring typical plasma parameters plus measurements of X-mode radiation from the $78-96$ GHz radiometer.]{Time histories of V-ECE measurement in runaway electron discharge $\#73002$ featuring typical plasma parameters plus measurements of X-mode radiation from the $78-96$ GHz radiometer.}
\label{fig:73002_51} 
\end{figure}

X-mode measurement in the frequency range $78-96$ GHz during  runaway electron discharge  $\# 73002$ on TCV is presented in Figure \ref{fig:73002_51}. A thermal plasma discharge is run after the runaway experiment, with the optimal diagnostic setup, and is used for calibration of the diagnostic. The generation of runaway electrons  is achieved by reducing the electron density at a plasma current $\sim 150$ kA. 
The runaway electron beam was generated at around $\sim 0.5$ s. The measurement shows a thermal phase before $\sim 0.5$ s. The energies at which the thermal contribution is detected are associated to the lower frequencies,  which have  their X$2$ cold resonances within the line of sight. The higher frequencies, which correspond to energies $\sim 153$ keV in Figure \ref{fig:73002_51}, have a smaller thermal contribution since their X$2$ thermal resonances are not within the line of sight. That is, from $\sim 0.5$ s ,  the runaway electron radiation is measured by the diagnostic  without thermal pollution from about $\sim 0.5$ s at some of the energies. Moreover, the drop in the plasma density and temperature should reduce the thermal emission from $0.5$ s at all energies. Finally, note that the flat ECE intensity after ~0.5s is not related to a saturation of the diagnostic. The runaway electron phase produces nearly constant intensity at V-ECE frequencies. 

\section{Analysis of the energy distribution}
\label{meas4c_disc_edf}

\subsection{Theory of V-ECE measured distributions}
\begin{figure}[ht!] 
\centering 
\includegraphics[width=.8\linewidth]{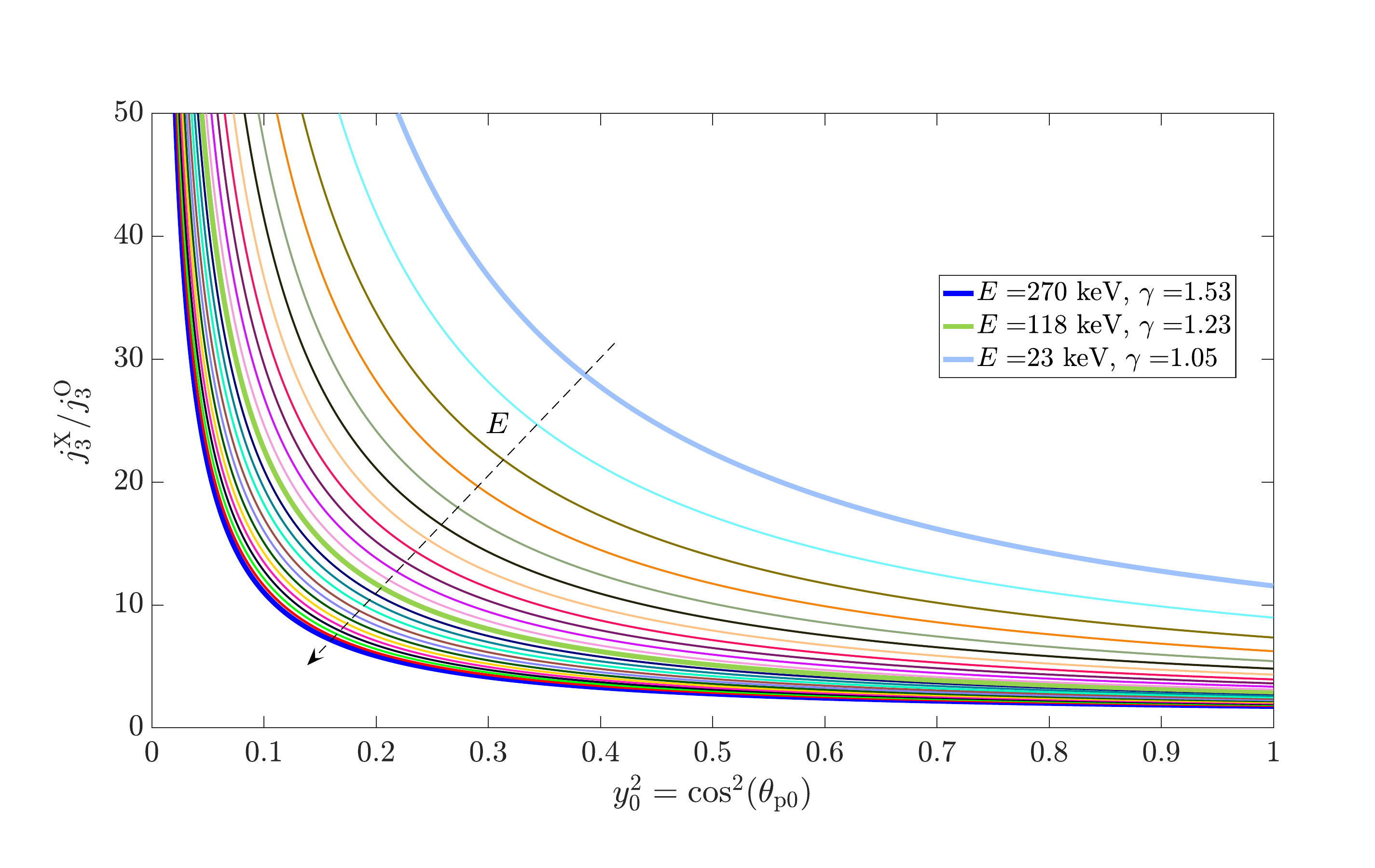} 
\caption[Ratio of X-mode to O-mode intensities as a function of pitch angle.]{Ratio of X-mode to O-mode intensities as a function of pitch angle at various energies.}
\label{fig:js_x_o_E} 
\end{figure}

By neglecting collective effects in the plasma, we aim to provide an expression of fast electron density at each energy (frequency) measured by V-ECE. The expression is found assuming a 2-parameters delta-type model distribution function,  which is appropriate for V-ECE that measures simultaneously 2 independent quantities (X- and O- mode power) at the same frequency. We start from the expressions of the single particle emissivities \cite{bekefi}, written for the O-mode as  
\begin{equation}
    \eta_{\mathrm{O}} =  \frac{e^2\omega^2}{8\pi^3\epsilon_0c} \sum_{n=1}^{\infty}\left[  \frac{p^2\cos^2\theta_{\mathrm{p}}}{1+p^2}  \mathcal{J}_n^2\left( \frac{\omega}{\omega_{\mathrm{ece}}}p\sin\theta_{\mathrm{p}} \right)  \right] 
    \label{eq7:eta_O}
\end{equation}
and for the X-mode as
\begin{equation}
    \eta_{\mathrm{X}} =  \frac{e^2\omega^2}{8\pi^3\epsilon_0c} \sum_{n=1}^{\infty}\left[  \frac{p^2\sin^2\theta_{\mathrm{p}}}{1+p^2}  \mathcal{J}_n^{'2}\left( \frac{\omega}{\omega_{\mathrm{ece}}}p\sin\theta_{\mathrm{p}} \right)  \right], 
    \label{eq7:eta_X}
\end{equation}
where $p$ is  the normalized momentum $p=\gamma m/c \overrightarrow{v}$ and  $\theta_{\mathrm{p}} $ the pitch angle defined such that $\mathrm{sin}(\theta_p) = v_{\parallel}/v$. 
$\mathcal{J}_n$ and $\mathcal{J}_n'$ representing respectively the Bessel function of the $n$ order and its derivative ($n$ still represents the harmonic number).  
 The emission coefficients of an uncorrelated ensemble of electrons is expressed as
\begin{equation}
    j_{\mathrm{O,X}} =  \int \eta_{\mathrm{O,X}}\mathcal{F}(\overrightarrow{p}) \mathrm{d}\overrightarrow{p}, 
    \label{eq7:jo}
\end{equation}
with $\mathcal{F}$ representing the distribution function. 
The emitted power is calculated performing spectral and spatial integrals of the particle emissivity as well as an integration over solid angle $\Omega_{\mathrm{S}}$. The spectral integration is performed in the frequency bandwidth around a given central frequency, $\omega$. The spatial integral is computed in the volume of plasma lying within the antenna pattern, estimated as the product of the effective area of the antenna $A$, in the radial direction, and the distance $s$, representing the  vertical extent of the antenna: 
\begin{equation}
    P_{\mathrm{O,X}} (\omega) =  \int  j_{\mathrm{O,X}}dsdAd\Omega_{\mathrm{S}}d\omega = \int \eta_{\mathrm{O,X}} \mathcal{F}(\overrightarrow{p}) \mathrm{d}\overrightarrow{p}dsdAd\Omega_{\mathrm{S}}d\omega.
    \label{eq7:po}
\end{equation}
Previous methods to analyze the energy distribution based on V-ECE polarisation measurements are reported in the literature \cite{Fidone_1990}. We use the method introduced in Reference \cite{Luce1988}, assuming a delta distribution of the form
\begin{equation}
    \mathcal{F}(p_0,y) = \frac{n_{\mathrm{fast}}}{2\pi p^2_0} \delta(p-p_0)\delta(y-y_0),
\end{equation}
where $y=\cos\theta_{\mathrm{p}}$ and $n_{\mathrm{fast}}$ is the number density of fast electrons, the ratio $ j_{\mathrm{X}} /  j_{\mathrm{O}}$ calculated in an appropriate coordinate system yields 

\begin{equation}
    \frac{j_{\mathrm{X}} (\omega)}{ j_{\mathrm{O}} (\omega)} = \frac{1-y_0^2}{y_0^2} \times \frac{\mathcal{J}_n'^2\big(\frac{\omega}{\omega_{\mathrm{ece}}}p_0\sqrt{1-y_0^2}\big)}{\mathcal{J}_n^2\big(\frac{\omega}{\omega_{\mathrm{ece}}}p_0\sqrt{1-y_0^2}\big)}.
    \label{eq:ratio}
\end{equation}
The ratio  determines directly a \textit{weighted} pitch angle, $y_0$. Since the \textit{weighted} pitch angle is obtained from the ratio of the emission coefficients prior to the integration, the inferred pitch angle and number density from measured power ratio represents an average over  bandwidth and over plasma volume. 
The normalized momentum $p_0$ is obtained directly from the measured frequency via the relation 
\begin{equation}
    p_0 = \sqrt{\Big(\frac{n \omega_{\mathrm{ece}}}{\omega}\Big)^2-1}.    
    \label{eq:p}
\end{equation}
Figure \ref{fig:js_x_o_E} shows the ratios of the X- to the O-mode intensities  computed using  Equation \ref{eq:ratio} for a range of energies $($normalized momentum$)$. The values of energies were found from the values of  $\omega_{\mathrm{ece}} / 10^{9}  \sim 2\pi \times 40$ rad/s ( $eB_{0}/m$ for $B_{0}=1.41$ T),  and selected frequencies $\omega$ in the range $\omega \sim 2\pi \times [96-114]$ GHz, corresponding to a V-ECE operational range. We note that the higher is the energy, the lower is the polarisation ratio than can be achieved. This means that there is a minimum energy required to achieve a  given ratio.  The form of the curves, near the high values of $y_0^2$, is relatively flat, meaning that an uncertainty in the ratio can yield a large uncertainty on the \textit{weighted} pitch angle. The assessment of the enhancement in the reconstructed distributions,  in one of the directions, parallel or perpendicular to the field,  is done by comparing the theoretical ratios in Figure \ref{fig:js_x_o_E} to the measured ratio of intensities. Comparisons with an isotropic distribution can be made at each energy, in the different scenarios to determine the direction of the enhancement of the distribution . At a fixed energy, the lower ratios would suggest an enhancement in the parallel direction while the higher ratios would suggest an enhancement in the perpendicular direction.

The number density of fast electrons is estimated from the the calibrated intensity measured in any of the polarisations. With X-polarisation, the expression of the emitted power within a frequency bandwidth is found computing the integral in Equation \ref{eq7:po}, which yields
\begin{equation}
    P_{\mathrm{BW,X}} = \frac{e^2c \widetilde{H_{\mathrm{p}}} \Delta f }{2\epsilon_0 \widetilde{f}} n_{\mathrm{fast}} p_0 (1-y_0^2) J_n'^2\big(\frac{\widetilde{f}}{f_{\mathrm{ece}}}p_0\sqrt{1-y_0^2}\big).
    \label{eq7:pbw}
\end{equation}
The terms $\widetilde{f}$ and $\widetilde{H_{\mathrm{p}}}$ represent respectively the mean frequency in the bandwidth  and the mean plasma height within the antenna pattern. The number density of electrons at energy 
\begin{equation}
    E = (\frac{nf_{\mathrm{ece}}}{\widetilde{f}}-1) m_{\mathrm{e0}}c^2, 
\end{equation}
measured from the $n$ harmonic, in the channel centered at the frequency $\widetilde{f}$ has the expression  
\begin{equation}
    n_{\mathrm{fast}} (\widetilde{f})  =  \frac{2\epsilon_0 \widetilde{f} P_{\mathrm{BW,X}}(\widetilde{f})}{e^2c \widetilde{H_{\mathrm{p}}} \Delta f p_0 (1-y_0^2) J_n'^2\big(\frac{\widetilde{f}}{f_{\mathrm{ece}}}p_0\sqrt{1-y_0^2}\big)}. 
    \label{eq:nfast}
\end{equation}

\subsection{Inferring the  electron energy distribution from V-ECE measurements}
\begin{figure}[ht!] 
\centering 
\includegraphics[scale=0.3]{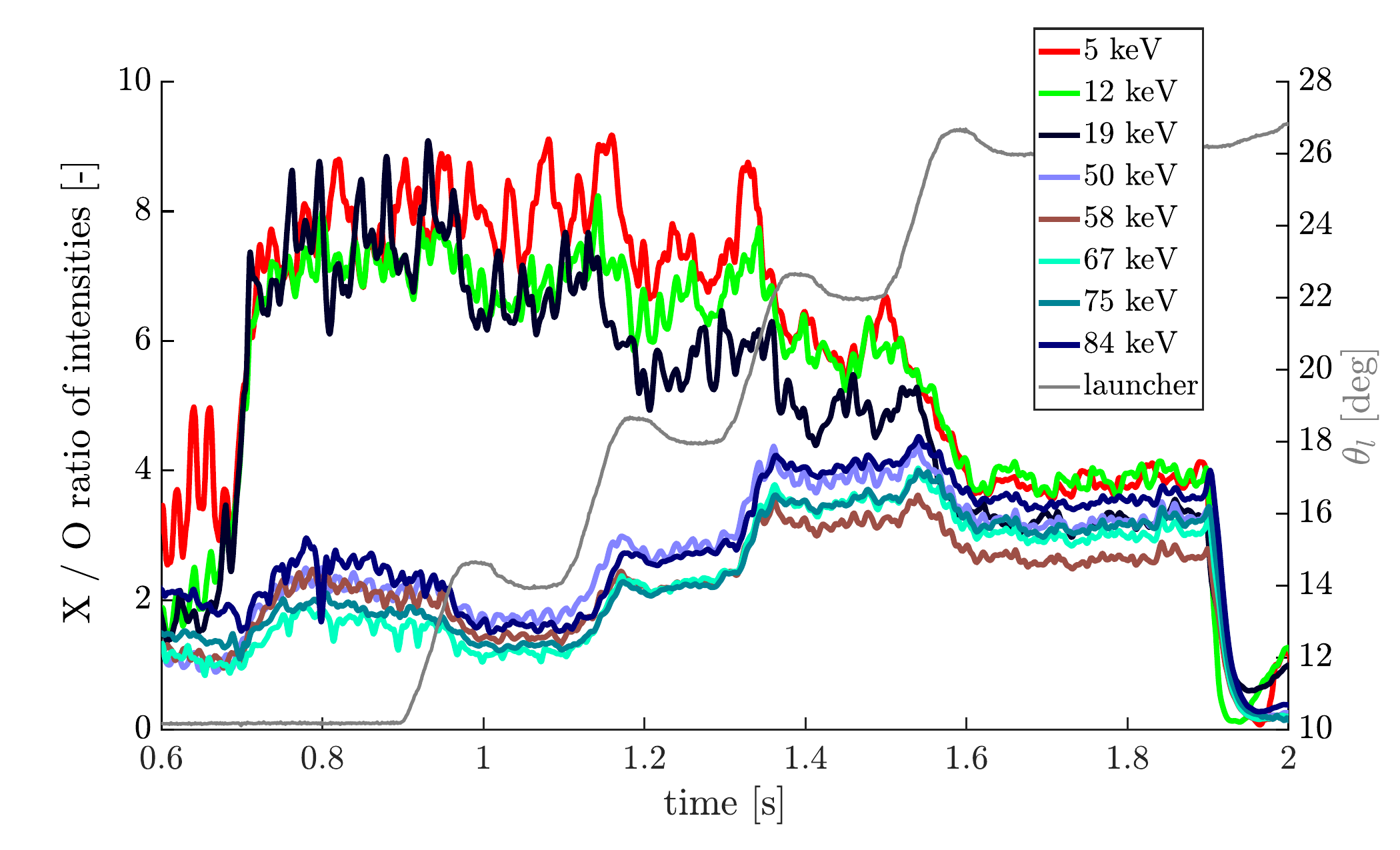} 
\caption[Ratio of polarisation during Current drive in TCV discharge $\#72644$]{polarisation ratio for Electron Cyclotron Current Drive discharge  $\#72644$ on TCV}
\label{fig:ratio_x_O} 
\end{figure}

The ratios of X- to O-mode intensities measured during the heating phase of the ECCD experiment $\#72644$ are shown in Figure \ref{fig:ratio_x_O}. The O-mode intensities were measured using the same radiometer, in the frequency range 96–114 GHz, with a twin discharge (an exact repeat of discharge $\#72644$), while the wire grid polariser was rotated 90° in the plane parallel to the grid. The energies are estimated, assuming harmonic overlap can be neglected, and that non-thermal contribution to the radiation comes essentially from the third harmonic emission. 

\begin{figure}[ht!] 
\centering 
\includegraphics[width=1.\linewidth]{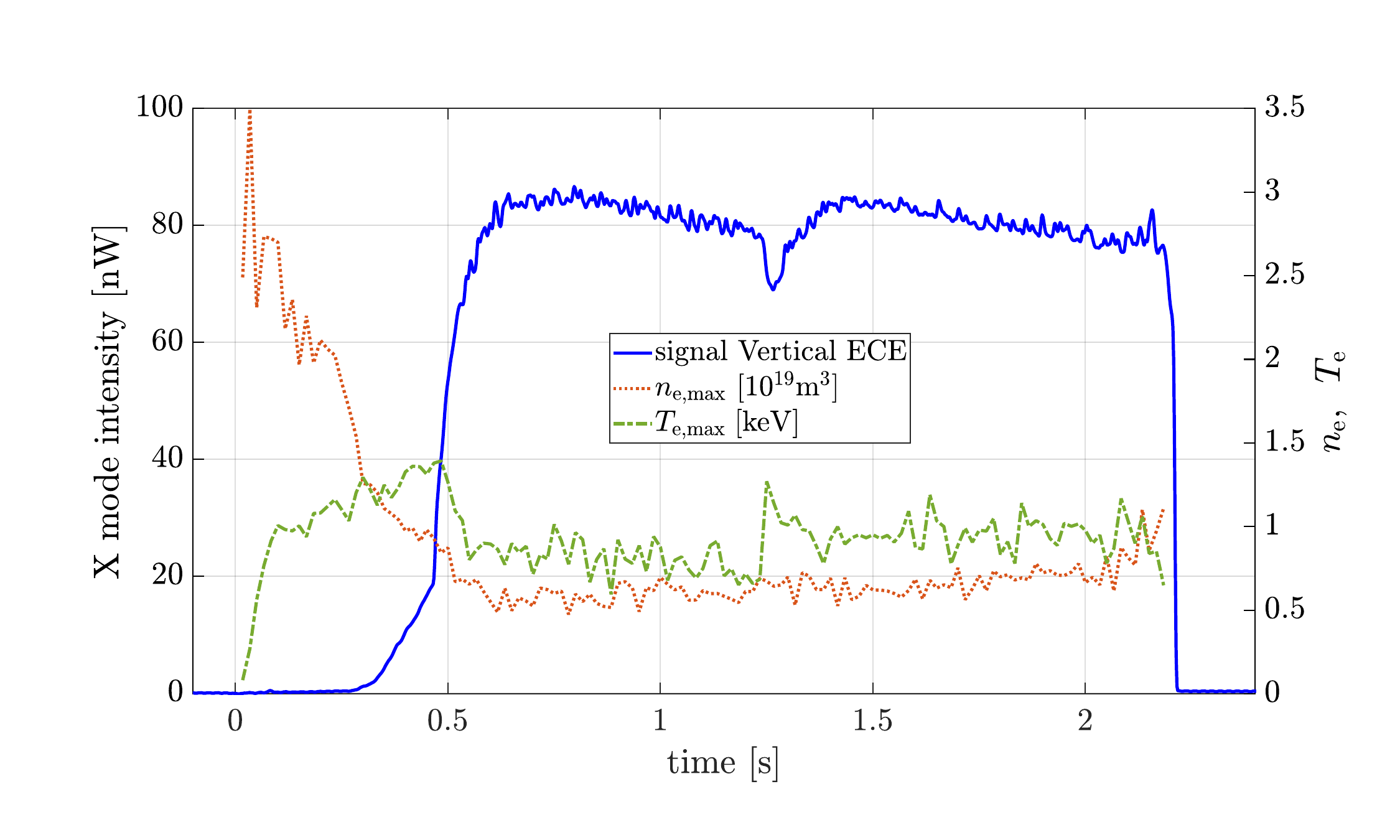} 
\caption[Runaway electron experiment used for the reconstruction of the non-thermal electron distribution.]{Runaway electron experiment used for the reconstruction of the non-thermal electron distribution. The Figure shows a calibrated V-ECE signal together with electron density and temperature. }
\label{fig:ex} 
\end{figure}

\begin{figure}[ht!] 
\centering 
\includegraphics[width=.7\linewidth]{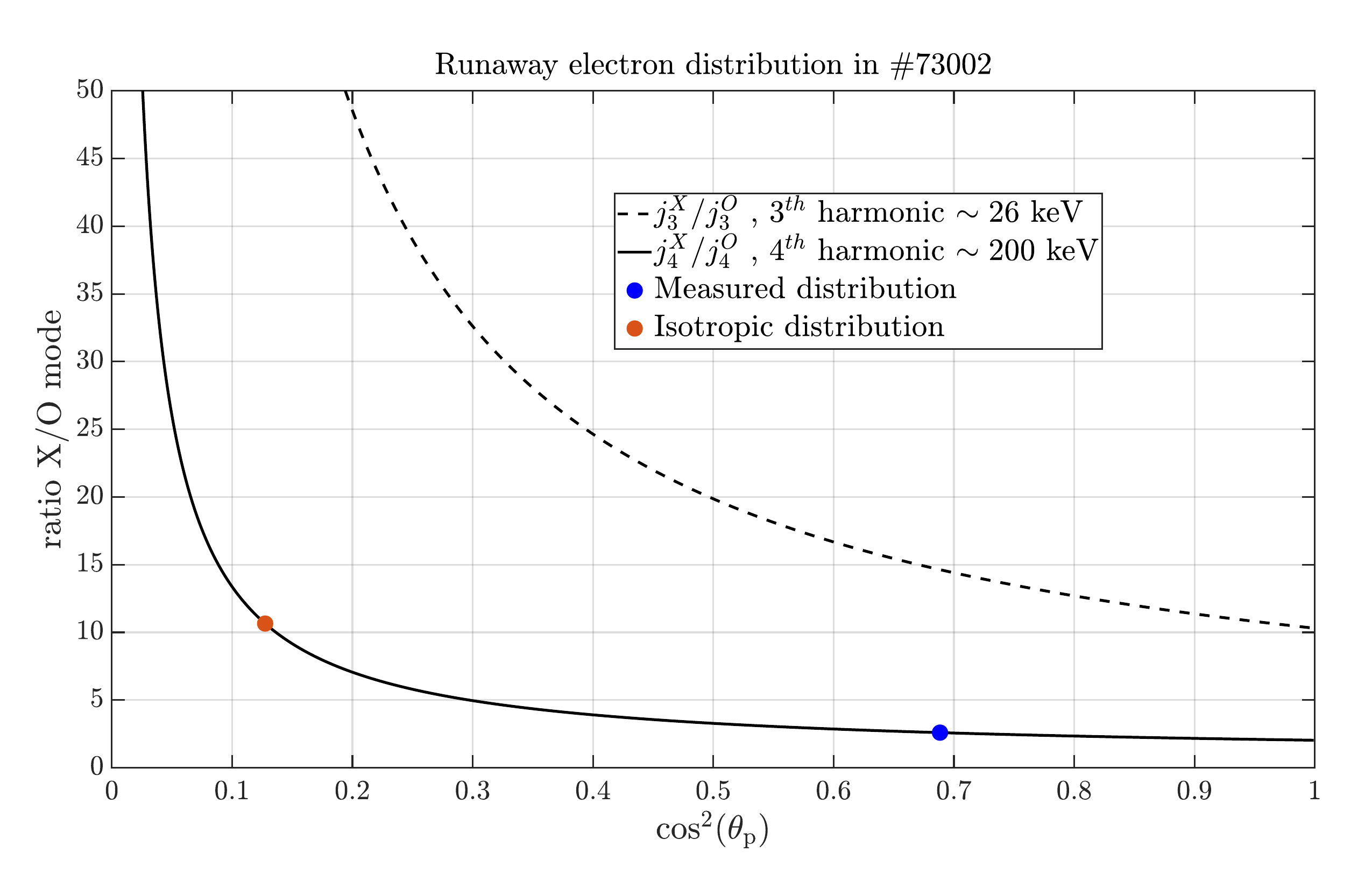} 
\caption[polarisation ratios of an isotropic distribution compared to the ratio from the runaway experiment.]{polarisation ratios of an isotropic distribution compared to the ratio from the runaway experiment.}
\label{fig:enhancement} 
\end{figure}

 For the lower energies, The ratio decreases  with increasing $\theta_L$, suggesting that the parallel energy is been preferentially enhanced after the first launcher steps. This could mean that the interaction between the lower energy electrons and the ECH wave, which in principle enhances perpendicular energy,  decreases with increasing  $\left| N_{\parallel} \right|$. In that case, the low energy electrons would have their energy enhanced from collisions with the higher energy electrons. The enhancement of the  parallel energy through collisions may be responsible for the reduction of X- to O-mode power ratio. At the lower energies, the ratio is higher for smaller $\left| N_{\parallel}\right| $ and launcher angle $\theta_L$. This trend is inverted at the higher energies, where the polarisation ratios have lower values in general. It is observed that small $\left| N_{\parallel}\right| $ leads to stronger enhancement in the perpendicular direction, explaining the higher ratio of the X- to the O- mode intensities. At higher $\left| N_{\parallel}\right| $ (approximately between $1.4$ s and $1.6$ s), the higher energies are more enhanced in the perpendicular direction, consistently with the expectations.  
 \begin{figure}[ht!] 
\centering 
\includegraphics[scale = 0.4]{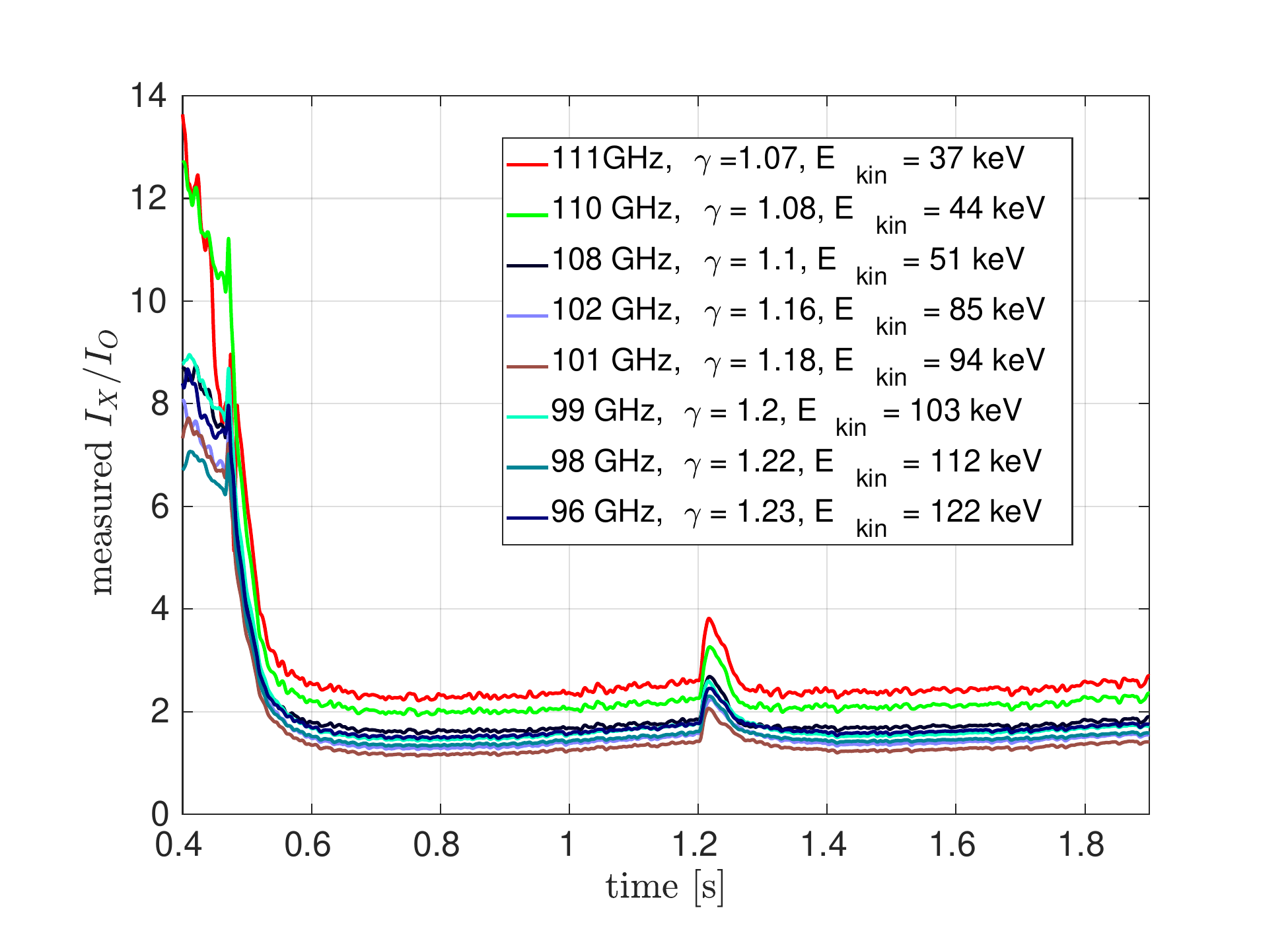} 
\caption[Ratio of the X-mode to the O-mode radiation Ratio of the X-mode to the O-mode radiation intensity for a runaway electron discharge]{Ratio of the X-mode to the O-mode radiation intensity for a runaway electron discharge.}
\label{fig:ratio} 
\end{figure}
The ratio of the intensities of the X- to the O-mode intensity for the runaway electron case $\# 73000$ is shown in Figure \ref{fig:ratio}. The ratio sharply decreases when the runaway beam starts, confirming that it is the parallel energy of the electrons that is been enhanced, in particular due to the toroidal electric field acceleration. 
In general the ratio is lower for runaway electron radiation, at each measured frequency, compared to the suprathermal electrons generated by ECCD  in the case presented in Figure \ref{fig:ratio_x_O}. Note that the discussed polarisation ratios are free from the calibration uncertainties mentioned in Section \ref{uncer}. The ratios are obtained from the radiometers’ raw data, accounting for the radiometers’ insertion losses. The insertion losses were determined by rotating the wire grid polariser by 45$^{\circ}$ with the plasma as a source,  and by using an external power source from a Vector Network Analyzer. \\
We focus on the runaway electron experiment $\# 73000$ on TCV for the reconstruction of the energy distribution. That is for two main reasons. The first reason is that the ECCD experiment is performed at a much higher density and necessitates the inclusion of the collective effects. The second reason is because the approach using the polarisation ratios may not be appropriate in the high density case, due to possible detection of radiation from multiple wall reflections, which will contaminates the ratio of polarisation when plasma refraction is not mitigated.
The polarisation ratios in the ECCD case can be useful for identifying enhancements in the parallel/perpendicular energy. However, the absolute values of these ratios in ECCD should not be directly compared to those in the runaway case. The analysis in the runaway electron case is based on Figure \ref{fig:ex}, where time traces  of V-ECE signal central density and temperature are shown. The time of the runaway beam formation is identified around $\sim 0.5$ s, and the ratio of intensities that is used is obtained after that time.  
The polarisation ratio for a single channel is shown on Figure \ref{fig:enhancement}. Assuming that the polarisation ratio obtained in the runaway case is not contaminated, a much higher energy is needed in order match the measured ratio, suggesting that the measured emission primarily comes from downshifted emission from the $4{\mathrm{th}}$ harmonic and not from the above $3^{\mathrm{rd}}$ harmonic.  With the emission coming from the $4^{\mathrm{th}}$ harmonic, the electron energy is sufficient to achieve the measured ratio of  polarised intensities. In general, the emission that is detected comes from more than one harmonic.  In the case of runaway electrons, the expected distribution is flat and,  higher energy emission can actually  dominate the total radiation. 
This analysis suggests a net enhancement in the parallel direction is observed  for runaway electrons as expected with a value of \textit{weighted} pitch angle close to $y_0^2 \sim 0.68 $.

\begin{figure}[ht!] 
\centering 
\includegraphics[width=.8\linewidth]{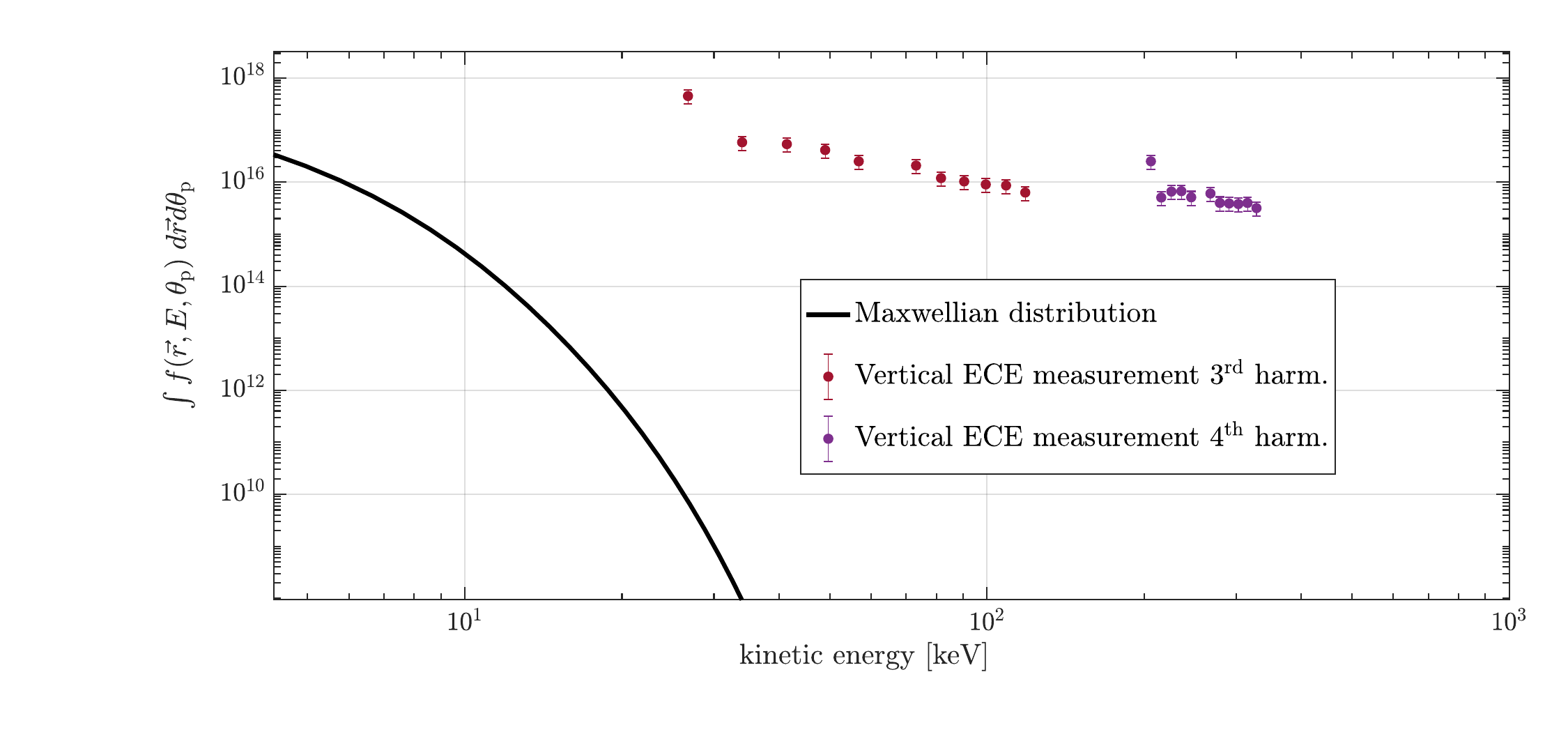} 
\caption[Example of reconstruction of the electron distribution in a runaway electron scenario .]{Example of reconstruction of the electron distribution in a runaway electron measurement with V-ECE using a 2-parameter model. }
\label{fig:edf_re} 
\end{figure}
We compute the density of runaway electrons at measured energy using Equation \ref{eq:nfast} and  the calibrated X-mode intensities ( shown for one channel on  Figure \ref{fig:ex}). We note that at the beginning of the discharge, before the runaway beam formation at $\sim 0.5$ s,  the measured power is nul and so is the number density of the runaway electrons. The calculated number density of runaway electrons at some energies, when the beam is formed,  is shown in Figure \ref{fig:edf_re}. The densities are compared to those of a Maxwellian distribution. We note that the numbers shown in the figure are the fast electron density found with Equation \ref{eq:nfast} and multiplied by a plasma volume, to obtain a total number of particle at a given energy. 
The  reconstructed energy distribution is shown for both the assumptions that the measured radiation comes primarily from downshifted  $3^{\mathrm{rd}}$ (lower energies) or downshifted  $4^{\mathrm{th}}$ harmonic (higher energy). The relatively flat tail,  and the large number of electrons present at that high energies in the reconstructed distribution, compared to a Maxwellian distribution, shows the possibility of using V-ECE measurement for for runaway electron distributions studies with. 
\section{Conclusion}
\label{concls}
This paper presented the first V-ECE measurements of non-thermal electrons on TCV. The diagnostic more easily discriminates the radiation according to the energy of the electrons, compared to horizontally-viewing ECE diagnostics.  The calibration of the diagnostic relied on the calculation of radiation intensity under the conditions of low optical thickness,  and was verified using the plasma black-body radiation. Uncertainties in the calibration was found to lie below 30$\%$. The main contributions to the error margin are the systematic errors in the Thomson Scattering  measurements of electron density and temperature profiles. Measurements of radiation from non-thermal electrons discharges were discussed. The measurements of linearly X- and O-polarised radiation were achieved in ECCD and runaway electron scenarios. The paper has described a scenario designed to combine  calibration and current drive in a single plasma discharge. In an ECCD scenario with varying launcher angle, stair-shaped V-ECE measurements and HXRS measurements allowed to observe how increasingly energetic electrons are excited with increasingly higher absolute values of ECH wave $N_{\parallel}$. From V-ECE measurements, it was observed an energy region, measured at around 104 GHz, showing a significant excitation by the ECH wave for values of $N_{\parallel} \sim -0.41$,$-0.51$, and $-0.61$. The observation suggested that electrons in that energy space preferentially receive the ECH wave energy due to the overall wave-plasma dynamics and not necessarily  to the direct damping of wave on those electrons. V-ECE measurements in runaway electron scenarios has facilitated the analysis of the electron energy distribution. Ratios of X- to O- mode radiation intensities used in a 2-parameters model for the energy distribution, has shown enhancement of the energy distribution in the parallel direction in runaway electron case, while enhancements in both parallel and perpendicular directions were found in ECCD case. Future work on TCV could include multi-diagnostic reconstruction of the  electron energy distribution  leveraging the advantages of V-ECE such as  great temporal resolution and one-to-one correspondence between energy and measured frequency in ideal conditions. 

The work described in this article displays important advancements over previously published results and has the potential to inspire future research on V-ECE diagnostics and calibration of ECE instruments. A key strength of this work is the interpretation of diagnostic results almost free from thermal background radiation. This achievement is made possible by better control over multiple wall reflections and represents an innovation compared to previous studies, where viewing dumps were considered \textit{sine qua non} conditions for operating the diagnostic. The control of background radiation, demonstrated for the first time in this work has notably enabled a more accurate cross-calibration of V-ECE with Thomson Scattering. Techniques that allow regular calibration of ECE diagnostics without requiring machine venting will be beneficial for next-generation machines such as ITER, SPARC, or DEMO. More broadly, the results presented in this article could help revive interest in fast electron diagnostics in Tokamaks using ECE. As demonstrated on TCV, V-ECE can play a key role in the detection and mitigation of low-energy runaway electrons, bringing us a step closer to achieving fusion energy.

\subsection{Acknowledgments}
The first author, A. T. B., thanks Olivier Sauter, Patrick Blanchard, Gerardo Giruzzi and Max Austin for their valuable insights. This work has been carried out within the framework of the EUROfusion Consortium, via the Euratom Research and Training Program (Grant Agreement No 101052200 — EUROfusion) and funded by the Swiss State Secretariat for Education, Research and Innovation (Staatssekretariat für Bildung, Forschung und Innovation — SBFI). Views and opinions expressed are however those of the author(s) only and do not necessarily reflect those of the European Union, the European Commission, or SBFI. Neither the European Union nor the European Commission nor SBFI can be held responsible for them. 
This work was supported in part by the Swiss National Science Foundation. 

\section*{References}
\bibliographystyle{unsrt}
\bibliography{biblio}

\end{document}